\documentclass[journal]{IEEEtran}
\ifCLASSINFOpdf
\else
\fi
\usepackage[T1]{fontenc}
\usepackage{textcomp}
\usepackage{xcolor}
\usepackage{amssymb,amsmath,amsfonts,amsthm}
\usepackage{mathrsfs}
\usepackage{array}
\usepackage{tabularx}
\usepackage{color}

\makeatletter
\let\MYcaption\@makecaption
\makeatother
\usepackage{subcaption}
\makeatletter
\let\@makecaption\MYcaption
\makeatother

\usepackage{balance}
\usepackage{mathtools}
\usepackage{tikz,pgf,pgfplots,pgfplotstable}
\usepgfplotslibrary{groupplots}
\usepgfplotslibrary{fillbetween}
\usetikzlibrary{spy, backgrounds}
\usetikzlibrary{patterns}
\usetikzlibrary{shapes.geometric, positioning}
\usepackage{tkz-euclide}

\tikzset{
    user/.pic = {     
        \draw[fill=white,rounded corners=0.02\R] (-0.09\R,-0.17\R) rectangle (0.09\R,0.17\R) ;                     
        \draw[fill=gray] (-0.08\R,-0.12\R) rectangle (0.08\R,0.12\R);                          
        \draw[rounded corners=0.005\R] (-0.04\R,0.14\R) rectangle (0.04\R,0.15\R);                     
        \draw (0,-0.14\R) circle (0.012\R) ; 
        \node (dim) at (0,0)  [align=center,minimum width=0.2\R,minimum height=0.3\R] {};
    }
}

\tikzset{
  server/.pic={
    \draw[fill=gray!20, draw=black, line width=0.6pt, rounded corners=2pt]
      (-0.4\R,-0.6\R) rectangle (0.4\R,0.6\R);
    
    \foreach \y in {-0.4\R, 0, 0.4\R} {
      \draw[fill=white] (-0.3\R,\y-0.1\R) rectangle (0.3\R,\y+0.1\R);
    }

    \foreach \y in {-0.4\R, 0, 0.4\R} {
      \foreach \x in {-0.2\R, 0, 0.2\R} {
        \fill[green!60!black] (\x,\y) circle (0.025\R);
      }
    }
  }
}

\tikzset{
  redserver/.pic={
    \draw[fill=gray!20, draw=black, line width=0.6pt, rounded corners=2pt]
      (-0.4\R,-0.6\R) rectangle (0.4\R,0.6\R);
    
    \foreach \y in {-0.4\R, 0, 0.4\R} {
      \draw[fill=white] (-0.3\R,\y-0.1\R) rectangle (0.3\R,\y+0.1\R);
    }

    \foreach \y in {-0.4\R, 0, 0.4\R} {
      \foreach \x in {-0.2\R, 0, 0.2\R} {
        \fill[red!60!black] (\x,\y) circle (0.025\R);
      }
    }
  }
}

\usetikzlibrary{calc,positioning,mindmap,trees,decorations.pathreplacing}
\usepackage{graphicx}
\usepackage{bbm}
\usepackage{bm}
\usepackage[utf8]{inputenc}
\usepackage{algorithm}
\usepackage{algpseudocode}
\usepackage{stfloats}
\usepackage[capitalize]{cleveref}
\usepackage{acronym}
\usepackage{siunitx}

\acrodef{ota}[OtA]{Over-the-Air}
\acrodef{mse}[MSE]{Mean Squared Error}
\acrodef{mmse}[MMSE]{Minimum Mean Squared Error}
\acrodef{snr}[SNR]{Signal-to-Noise-Ratio}
\acrodef{dp}[DP]{Differential Privacy}

\newtheorem{theorem}{Theorem}
\newtheorem{corollary}{Corollary}

\newtheorem{definition}{Definition}

\newtheorem{remark}{Remark}
\newtheorem{example}{Example}

\usepackage{soul}

\IEEEoverridecommandlockouts
\ifCLASSINFOpdf
\else
\fi
\newcommand{\norm}[1]{\left\lVert #1 \right\rVert}
\newcommand{\conj}{^*}
\newcommand{\abs}[1]{\left|#1\right|}
\newcommand{\CN}[2]{\mathcal{CN}\left( #1,#2 \right)}
\newcommand{\N}[2]{\mathcal{N}\left( #1,#2 \right)}
\newcommand{\reals}{\mathbb{R}}
\newcommand{\complexes}{\mathbb{C}}
\newcommand{\vect}[1]{\textbf{#1}}
\newcommand{\EX}[2][]{\mathbb{E}_{#1}\left\{#2\right\}}
\newcommand{\Var}[1]{\textnormal{Var}\{#1\}}
\newcommand{\Cov}[1]{\textnormal{Cov}\left[ #1 \right]}
\newcommand{\argmax}{\mathop{\rm arg~max}\limits}
\newcommand{\argmin}{\mathop{\rm arg~min}\limits}

\IEEEpubid{%
  \parbox{\textwidth}{\centering\footnotesize
    \vspace*{1.0cm}
    \copyright~2026 IEEE. Personal use of this material is permitted.  Permission from IEEE must be obtained for all other uses, in any current or future media, including reprinting/republishing this material for advertising or promotional purposes, creating new collective works, for resale or redistribution to servers or lists, or reuse of any copyrighted component of this work in other works.
  }%
}

\begin{document}

\title{Secure Over-the-Air Computation Against Multiple Eavesdroppers using Correlated Artificial Noise}

\author{
\thanks{This work was supported in part by Zenith, ELLIIT, Swedish Research Council (VR), Knut och Alice Wallenbergs stiftelse (KAW), Wallenberg AI, Autonomous Systems and Software Program (WASP) funded by the Knut and Alice Wallenberg Foundation, and by the Bavarian Ministry of Economic Affairs, Regional Development and Energy within the scope of the 6G Future Lab Bavaria.}
David~Nordlund,~\IEEEmembership{Student Member,~IEEE,}
        Luis~Maßny,~\IEEEmembership{Student Member,~IEEE,}
    Antonia~Wachter-Zeh,~\IEEEmembership{Senior Member,~IEEE,}
    Erik~G.~Larsson,~\IEEEmembership{Fellow,~IEEE,}
    and~Zheng~Chen,~\IEEEmembership{Senior Member,~IEEE}%
    \thanks{D. Nordlund, E. G. Larsson and Z. Chen are with Linköping University, Dept. of Electrical Engineering (ISY), Division of Communication Systems, 581 83 Linköping, Sweden (emails: david.nordlund@liu.se, erik.g.larsson@liu.se, zheng.chen@liu.se).}%
    \thanks{L. Maßny and A. Wachter-Zeh are with the School of Computation, Information and Technology at the Technical University of Munich, 80333 Munich, Germany (emails: luis.massny@tum.de, antonia.wachter-zeh@tum.de).}
    \thanks{Part of this work appeared in a preliminary study presented at the 2023 IEEE Information Theory Workshop (ITW)~\cite{massny2023secure}.}
}

\maketitle

\begin{abstract}
Over-the-air (OtA) computation enables scalable analog aggregation by exploiting the superposition property of wireless channels, making it an attractive joint communication and computation paradigm for distributed sensing and learning. However, the uncoded nature of analog transmission exposes the computation result to eavesdropping, and the fundamental security limits of OtA computation against multiple cooperating adversaries remain poorly understood. In this paper, we develop an estimation-theoretic framework for analyzing the security of analog OtA computation in the presence of multiple distributed eavesdroppers that may jointly process their observations. We first derive the optimal estimator for cooperating eavesdroppers and bounds on the achievable estimation accuracy of both the legitimate receiver and the adversaries. Our analysis reveals a key insight: while random channel phase misalignment provides significant inherent MSE-security against individual eavesdroppers, this protection largely disappears once multiple eavesdroppers cooperate. Motivated by this observation, we propose a correlated artificial noise design based on zero-forcing that preserves the aggregation accuracy at the legitimate receiver while maximizing the estimation error at the cooperative eavesdroppers. Numerical results demonstrate that the proposed design substantially reduces the security advantage gained through eavesdropper cooperation and achieves security close to uncorrelated artificial-noise schemes without sacrificing computation accuracy. These results provide both a theoretical characterization of the security limits of analog OtA computation and a practical design guideline for its secure deployment in real-world wireless systems.
\end{abstract}

\begin{IEEEkeywords}
Over-the-air computation, secure communication, correlated artificial noise, optimization.
\end{IEEEkeywords}

\section{Introduction}
Recent advances in artificial intelligence (AI) are reshaping the design of next-generation communication networks. Emerging decentralized AI applications, such as federated learning (FL)~\cite{mcmahanCommunicationEfficientLearningDeep2017}, rely on iterative information exchange between a large numbers of distributed devices and a central server. However, this process can create significant communication bottlenecks, particularly in wireless systems where the exchange of high-dimensional model updates is constrained by limited resources, such as time–frequency resource blocks and power budgets.
\ac{ota} computation addresses this issue by exploiting the multiple access channel as a natural aggregation operator~\cite{otaNazer,
  otaGoldenbaum,otaZhu,
  ota-old-new,Wang-ota6G,Chen-ota-multiple-functions, Cao-ota}, enabling direct function computation from distributed sources without orthogonal resource allocation.
\cite{ota_survey_nomographic}. Consequently, the system is innately
scalable in the number of transmitting nodes. Among its many promising applications in distributed systems, \ac{ota} FL has been extensively studied in recent works~\cite{YangotaFL, Cao-otaFL,
  zhuBroadbandAnalogAggregation2020,
  amiriFederatedLearningWireless2020, abrarBiasOverTheAir2024}.

Research in \ac{ota} computation has grown considerably in recent years, spanning many variants depending on the nature of information sources (analog or digital), data processing designs (coded or uncoded), and the availability of channel state information (coherent or non-coherent combining). 
In this work, we consider analog \ac{ota} computation with linear precoding and coherent processing. A large portion of the literature focuses on optimizing the processing functions, whereas the security and privacy aspects have received relatively less attention. In particular, the limited use of digital encryption and advanced coding schemes renders \ac{ota} computation systems vulnerable to both jamming and eavesdropping attacks.
With jamming attacks, the goal is to disturb the system and corrupt the computation results.
The countermeasures have been studied in both pure \ac{ota} computation and \ac{ota}-FL settings, including detection~\cite{nordlund2023detecting, adhikary2025jamming, 
adhikary2025denial} and mitigation strategies~\cite{adhikary2025jamming, adhikary2025denial, 
ezawa2025mitigating}, as well as the characterization of optimal attack mechanisms~\cite{wang2025byzantine}.
In this work, we focus on privacy threats, where an external eavesdropper can overhear the transmitted signals and estimate the computation result. Traditional information-theoretic security measures are not directly applicable to the \ac{ota} setting, where the objective is to recover an aggregated function of the data rather than infer individual data distributions.
For this reason, we adopt the notion of \textit{estimation-theoretic security}, which aims at preventing an eavesdropper from accurately estimating a secret~\cite{guoEstimationWSN2017,gokenEstimationTheoreticEncoding2019}, measured by the eavesdropper's \ac{mse}.
Perfect information-theoretic secrecy (zero mutual information) implies that an eavesdropper's observation is statistically independent from the secret.
In contrast, having a maximal \ac{mse} implies that an eavesdropper cannot estimate the secret more accurately (in the \ac{mse} sense) than from its prior knowledge; however, this does not necessarily induce statistical independence.

\newpage
Common approaches to estimation-theoretic
security includes the design of encoding and decoding functions~\cite{reboredoFilterDesignSecrecy2013,gong2025joint}, cooperative
jamming~\cite{yangCooperativeJamming2013}, cooperative precoding~\cite{ozcelikkaleCooperativePrecording2015}, and
beamforming~\cite{peiMaskedBeamforming2012,wang2025secure,wang2026active}.
Another popular idea is to leverage artificial
noise to interfere with the eavesdropper~\cite{goelGuaranteeingSecrecyUsing2008,swindlehurstFixedSINR2009,khistiGaussianMIMOWiretap2007, privateMMSE,mao2025joint}. 
Particularly, artificial noise has been extensively studied in the context of differentially private~\cite{dwork2014algorithmic} FL with \ac{ota} computation, in both single-antenna~\cite{OtAFLDP, privacy-for-free,FLdithering, taoOtAFLDP, park2023differential} and multiple-antenna systems~\cite{liuDPOtaFLMIMO}. In this scenario, noise is injected to protect local gradients from revealing information about the training dataset~\cite{xiaoOtAFLoverview}.
In our work, we aim at protecting the intended \ac{ota}
computation result against eavesdroppers while ensuring a high level of aggregation accuracy at the legitimate receiver. Following the estimation-theoretical approach in prior works \cite{ota_survey_nomographic,privateMMSE,He2024secure,reboredoFilterDesignSecrecy2013}, we measure accuracy and security in terms of the \ac{mse} at the legitimate receiver and at the eavesdroppers, respectively.

\subsection{Closely Related Works}
\label{sec:related-works}
Prior work on secure \ac{ota} computation has mainly focused on two approaches: 1) \textit{cooperative jamming} and 2) \textit{artificial noise injection}.
In \cite{friendly-jammer}, a friendly jammer, whom the server can perfectly compensate for, is leveraged to degrade the eavesdropper's estimation performance. 
The receiver itself can act as a friendly jammer when full-duplex receivers are available \cite{server-jammer, luo2023joint, yao2025secure,li2023joint}, as it can perfectly cancel the jamming signal locally. Alternatively, \cite{yan2025secure}
considers selecting a subset of users to act as jammers, with the drawback of excluding some users in the computation process. 
In \cite{goldenbaumSecureComputation2016}, a
joint source-channel coding scheme is proposed for
information-theoretic security. However, the analysis is limited to linear discrete channels. A cryptographic approach for fading-less channels was pursued in~\cite{bruneMultiParty2023}.

Following the artificial noise approach, earlier studies primarily focus on uncorrelated designs, where independent random Gaussian noise is added to the transmitted signals to obfuscate the source data symbols \cite{OtAFLDP,privacy-for-free}. A key drawback of this approach is that the injected noise directly affects the computation result, leading to degraded aggregation accuracy at the receiver. More recently, several works have explored the use of correlated noise to better balance the utility–privacy tradeoff \cite{massny2023secure,liaoOvertheAirFederatedLearning2022,He2024secure}.
In particular, adding zero-forced noise across multiple users enables noise cancellation at the receiver, thereby achieving privacy protection without degrading aggregation accuracy~\cite{massny2023secure, liaoOvertheAirFederatedLearning2022}.
Additionally, \cite{He2024secure} considers a full-duplex adversary acting as both eavesdropper and jammer and proposes a numerical algorithm to jointly design power allocation, user selection, artificial noise, and receiver filters. The approach of correlated artificial noise can be interpreted as joint beamforming across distributed antennas. A similar concept is considered in \cite{Iqbal2024secure}, which introduces a distributed null-steering beamforming scheme aimed at maximizing the \ac{mse} 
at the eavesdropper, although it relies on knowledge of the eavesdropper’s location.

\subsection{Novelty and Contributions}
We extend the study of estimation-theoretic security via correlated artificial noise in the scenario of multiple cooperating eavesdroppers. Building on the framework used for real-valued channels and a single-antenna eavesdropper in \cite{massny2023secure}, this work investigates the effects of multiple cooperative eavesdroppers and complex-valued channels, taking into account random phase shifts at the eavesdroppers. A more comprehensive experimental evaluation provides deeper insights into the impacts of system parameters and tradeoffs.
Our specific contributions are as follows.
\begin{itemize}
    \item Assuming Gaussian distributed source data, we provide
      \textit{bounds on the expected achievable \ac{mse}} at the
      eavesdroppers and at the server, in both cooperative and
      non-cooperative (eavesdropping) cases.

    \item Furthermore, we provide \textit{an optimal joint estimator for cooperative eavesdroppers} when they have exact channel state information. In an alternative case where they only have statistical information about the channels, and the phase follows a uniform distribution, we show that the system achieves maximum security level from an estimation-theoretic perspective, even without adding extra artificial noise.
      
    \item We illustrate the inherent security against eavesdropping in
      \ac{ota} systems, due to channel phase misalignment. Subsequently, we
      show that this security is compromised in the presence of
      cooperating eavesdroppers, motivating the need for deliberate
      security measures.
      
    \item Lastly, we propose an \textit{optimized artificial noise design} for the non-cooperative case, which also demonstrates strong security performance in the cooperative case, as shown in numerical simulations.
\end{itemize}

\subsection{Paper Organization and Notation}
The rest of the paper is organized as follows. Section
\ref{sec:system_model} introduces the security and accuracy metrics, provides an overview of the system, and discusses the power control design
and performance measures of the legitimate receiver. Section \ref{sec:security_level} provides the security bound and optimal strategy of the eavesdroppers. In Section \ref{sec:optnoise}, we
detail our proposed artificial noise design. Simulation results are presented in Section \ref{sec:results}, and the work is summarized in
Section \ref{sec:conclusion}.

Vectors and matrices are denoted by bold lower case and upper case symbols, respectively. All vectors are column vectors. $(\cdot) ^H$ denotes complex conjugate transpose, $(\cdot)\conj$ complex conjugate, $\norm{\cdot}$ the $\ell^2$ norm and $\CN{\cdot}{\cdot}$ the complex Gaussian distribution.

\section{System Model}
\label{sec:system_model}

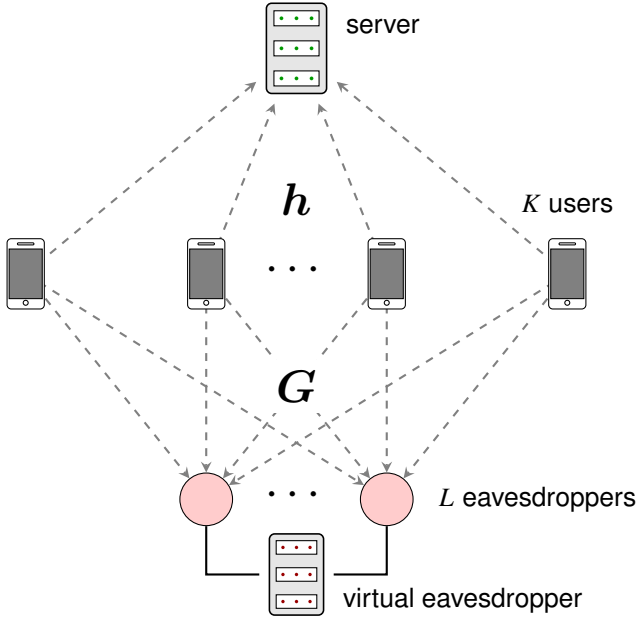
\begin{figure}
\centering
\scalebox{0.8}{%
\begin{tikzpicture}[>=stealth]

    \tikzset{
    usernode/.style={draw, circle, minimum size=1.2cm, fill=cyan!20},
    eavnode/.style={draw, circle, minimum size=0.7cm, fill=red!20}}
    
    \coordinate (center) at (0,0);

    \node[anchor=center] at (0, 4.5) (server) {
        \begin{tikzpicture}
            \newdimen\R
            \R=1cm
            \draw (4,-0.5) pic (dlap) {server} ;
        \end{tikzpicture}
    };
    \node[anchor=west] at ($(server.east) + (0,+0.3cm)$) {server};

    \coordinate (user1) at (-3.6,1.5);
    \coordinate (user2) at (-1.2,1.5);
    \coordinate (user3) at (1.2,1.5);
    \coordinate (user4) at (3.6,1.5);
 
    \node at (0,1.5) {\Huge $\cdots$};

    \node[anchor=south] at ($(user4.east) + (0,0.7cm)$) {$K$ users};
    
    \foreach \u in {user1, user2, user3, user4} {
    \draw[->, thick, dashed,gray] (\u) -- (server);
    }

    \node[fill=white, text=black, inner sep=1pt] at (0,2.5) {\huge $\bm{h}$};

    \node[eavnode] at (-1.2,-1.5) (eav1) {};
    \node[eavnode] at (1.2,-1.5) (eav2) {};
    
    \node at (0,-1.5) {\Huge $\cdots$};
    
    \node[anchor=west] at ($(eav2.east) + (0.2cm,0)$) {$L$ eavesdroppers};
    
    \foreach \u in {user1, user2, user3, user4} {
    \foreach \e in {eav1, eav2} {
      \draw[->, thick,dashed,gray] (\u) -- (\e);
    }
    }
    \node[fill=white, text=black, inner sep=4pt] at (0,0) {\huge $\bm{G}$};
    
    \newdimen\R
    \R=2.75cm
    \draw (user1) pic (dlap) {user} ;
    \draw (user2) pic (dlap) {user} ;
    \draw (user3) pic (dlap) {user} ;
    \draw (user4) pic (dlap) {user} ;

    \node[anchor=center] at (0, -2.5) (server2) {
        \begin{tikzpicture}
            \newdimen\R
            \R=0.9cm
            \draw (4,-0.5) pic (dlap) {redserver};
        \end{tikzpicture}
    };
    \node[anchor=west] at ($(server2.east) + (0, -0.35cm)$) {virtual eavesdropper};

    \foreach \eav in {eav1, eav2} {
        \draw[thick] (\eav) |- (server2);}
    
\end{tikzpicture}%
}  
\caption{Overview of the system. Wireless transmissions (dashed arrows), intended for the server, are overheard by $L$ eavesdroppers that jointly process their received signals at a \textit{virtual eavesdropper}. The solid lines represent an error-free channel.}
\label{fig:system_sketch}
\end{figure}

\renewcommand{\arraystretch}{1.5}
\begin{table}
\begin{center}
\caption{Notation}
\label{tab:notation}
\begin{tabular}{| c | c |}
\hline
 $K$ & number of legitimate users \\
\hline
 $L$ & number of eavesdroppers \\
\hline
$\bm{\gamma} = (\gamma_1,\ldots,\gamma_K)^T \in \complexes^{K\times 1}$ & source data \\
\hline 
 $s=\sum_{k=1}^K \gamma_k \in \complexes$ & the target function \\
\hline
 $\vect{h} \in \complexes^{K\times 1}$ & server channel vector \\
\hline
 $\bm{G}\in \complexes^{L\times K}$ & eavesdroppers channel matrix \\
\hline
 $y \in \complexes$ & received value at the server \\
\hline
 $z_\ell \in \complexes$ & received value at the $\ell$-th eavesdropper \\
\hline
 $D \in [0,1]$& approximation error \\
\hline
 $S \in [0,1]$ & security level \\
\hline
 $\bm{A}\in \complexes^{K\times m}$ & artificial noise precoding matrix \\
\hline
 $\vect{w} \in \complexes^{K\times 1}$ & artificial noise vector \\
\hline
 $\eta \in \reals$ & amplitude scaling factor \\
\hline
 $\vect{p} \in \complexes^{L\times 1}$ & eavesdropper combining vector \\
\hline
\end{tabular}
\end{center}
\end{table}
We consider a wireless system in which local data from $K$ distributed users are aggregated at a central server via \ac{ota} computation over a multiple-access block-fading channel. The system also contains $L$ passive eavesdroppers that attempt to estimate the intended computation result from their received signals, as illustrated in Fig.~\ref{fig:system_sketch}.\footnote{We
distinguish the problem of protecting the computation result from the
problem of protecting the input data, even though solutions to the latter problem would be applicable to the former.}  Let $\bm{\gamma} = \left( \gamma_1, \ldots , \gamma_K\right)
\in \complexes^K$ represent the source data vector, where $\gamma_k \in \complexes$
is the source data of user $k\in
\{1,\ldots,K\}$.\footnote{If the original data is real-valued, it can be
split into two streams that are then combined into one complex
stream.} 
We assume $\bm{\gamma} \sim
\CN{\bm{0}}{\bm{I}_K}$, meaning that the source data from all users follow independent Gaussian distributions and are normalized prior to transmission.%
\footnote{
The Gaussian assumption allows us to give results in closed forms, thus providing additional insights. Moreover, a Gaussian source states a lower bound regarding the server's estimation capabilities, cf.~\cite[Thm 8.6.6]{coverElementsOfInformationTheory}.
}
The goal is to compute the
function $s=f(\gamma_1,\ldots,\gamma_K) = \sum_{k=1}^K \gamma_k$. Note
that this framework can easily be extended to any weighted sum instead of the sum. 

The joint communication-computation processing chain can be divided into three main parts:
\begin{enumerate}
    \item encoding/pre-processing data at the source nodes;
    \item amplitude modulation and signal transmission over analog channels;
    \item decoding/post-processing of received superimposed signal at the receiver/server.
\end{enumerate}
A typical encoding/pre-processing rule in \ac{ota} computation is to perform channel inversion. 
Let $x_k \in \complexes$ represent the encoded data symbol from user $k$, which can be written as
\begin{equation}
     x_k = \eta h_k^{-1}\gamma_k,
\end{equation}
where $h_k\in \complexes$ is the channel fading coefficient between user $k$ and the server, and $\eta$ is an (amplitude) scaling factor chosen to satisfy the power constraints
$\EX{\abs{x_k}^2} \leq P$. 

The signals transmitted from legitimate users are received by both the server and the eavesdroppers.
The server receives
\begin{equation}
    y = \vect{h}^T \vect{x} + n_y,
\end{equation}
where $\vect{x}=(x_1,\ldots,x_K)^T$, $n_y \sim \CN{0}{\sigma_y^2}$ is additive noise and $\vect{h}=(h_1,\ldots,h_K)^T \in \complexes^K$ contains the corresponding channel coefficients. 
After receiving the superimposed signal, the server performs linear scaling (post-processing) and obtains an estimate of the intended computation result as 
\begin{equation}
    \hat{f}(y)=\frac{y}{\eta}=\sum_{k=1}^K \gamma_k+ \frac{n_y}{\eta}.
\end{equation}
The common scaling factor $\eta$ will determine the estimation noise in the computation result. 

At the other side, each eavesdropper $\ell\in \{1,\ldots,L\}$ receives
\begin{equation}
    z_\ell = \vect{g}_\ell^T \vect{x} + n_{z_\ell},
\end{equation}
where $n_{z_\ell} \sim \CN{0}{\sigma_z^2}$ is additive noise, and  $\vect{g}_\ell \in \complexes^K$ is the channel coefficient vector between the $K$ legitimate users and the $\ell$-th eavesdropper. The purpose of this work is to investigate the level of secrecy and aggregation accuracy of the system, depending on the strategy employed by the eavesdroppers. Although we consider only single-antenna devices, multiple single-antenna eavesdroppers can be viewed as a virtual eavesdropper with multiple distributed antennas. In particular, we investigate a scenario where the eavesdroppers can jointly post-process their received signals (using an error-free side channel) to obtain an estimate of the sum $s=\sum_{k=1}^K \gamma_k$. Table \ref{tab:notation} provides a summary of the notation used throughout this paper.

\subsection{Accuracy and Security Measures}
We adopt the following \ac{mse}-based accuracy and security measures.

\begin{definition}
    \label{def:security}
    Let $f: \complexes^K \to \complexes$ be the objective function. 
    For a given channel realization, we say that an \ac{ota} computation scheme, operating in the presence of multiple cooperating eavesdroppers, is \ac{mse}-approximate with approximation error
    $$D =  \min_{d_y: \complexes \to \complexes} \frac{\EX{ \vert d_y(y;\bm{\Phi}) - f(\gamma_1,\gamma_2,\ldots,\gamma_K) \vert ^2 \vert \bm{\Phi}}}{\Var{f(\gamma_1,\gamma_2,\ldots,\gamma_K)}} , $$
    and \ac{mse}-secure with security level 
    $$S=\min_{d_z: \complexes^L \to \complexes} \frac{\EX{\vert d_z(z_1,\ldots,z_L ; \bm{\Phi}) - f(\gamma_1,\gamma_2,\ldots,\gamma_K) \vert ^2 \vert \bm{\Phi}}}{\Var{f(\gamma_1,\gamma_2,\ldots,\gamma_K)}},$$
    where $d_y(\cdot ;\bm{\Phi})$ and $d_z(\cdot ;\bm{\Phi})$  are the post-processing functions of the legitimate server and eavesdroppers (jointly) respectively, and $\bm{\Phi} = \{ \Phi_1 , \ldots , \Phi_L\}$ is some side information about the channels between legitimate users and eavesdroppers. The expectations are taken over the joint distribution of $\bm{\gamma}$, the randomness used in the pre-processing, the fading distribution (if not part of the side information), and the channel noise $n_y$ or $(n_{z_1},\ldots,n_{z_L})$, respectively.
\end{definition}
According to this definition, the approximation error is the minimum (normalized) \ac{mse} that the server can achieve (given optimal post-processing), and the security level is a lower bound on the (normalized) \ac{mse} that the eavesdroppers can achieve after joint processing, reflecting the maximum information leakage.
The approximation error and security level lie in the range $D,S \in [0,1]$, since normalized \ac{mse} with maximum value $1$ can be achieved with the naive estimates $d_y(y;\bm{\Phi}) = \EX{s}$ and $d_z(z_1,\ldots,z_L ; \bm{\Phi}) = \EX{s}$ respectively. However, we note that a maximal estimation-theoretic security level of $S=1$ does not necessarily mean that no information is leaked about the source data from the received signal, as demonstrated in the following example.

\begin{example}
Consider a scalar random variable $X$ with distribution symmetric around $0$, e.g., $X \sim \N{0}{1}$ and suppose an eavesdropper observes $Y=|X|$. Then $\EX{X|Y=y} = \frac{1}{2} \cdot y + \frac{1}{2} \cdot (-y) = 0 = \EX{X}$. That is, the adversary cannot estimate $X$ more accurately (in terms of the \ac{mse}) than it could from the prior distribution, giving an \ac{mse} security level of $S=1$. Nonetheless, the observation $Y$ reveals $I(X;Y) = h(X) - h(X|Y) = \frac{1}{2} \log_2(2 \pi e) - 1 \approx \qty{1.05}{\bit}$ of information about $X$ to the eavesdropper.
\end{example}

\begin{remark}
    The \ac{mse} security considered in our work can be related to \ac{dp} via the noise variance. Their relation is most explicit when a secret $\bm{s}$ is perturbed by additive Gaussian noise as $\bm{z} = \bm{s} + \N{\bm{0}}{\sigma^2 \bm{I}_K}$:
    \begin{enumerate}
        \item
        When the secret data is clipped with the range $[-C,+C]$, then the output satisfies $(\epsilon,\delta)$-\ac{dp} with $\epsilon^2 = 2\ln(1.25/\delta) \frac{ C^2 }{ \sigma^2 }$~\cite{dwork2014algorithmic}. A larger variance $\sigma^2$ yields stronger privacy (smaller $\epsilon$).
        This is the core mechanism of the \ac{dp}-SGD implementation~\cite{abadi2016dpsgd}, which is widely adopted for single-client machine learning settings.
        In federated learning with multiple clients, a similar relationship holds, although the noise variance depends additionally on the \ac{ota} power control coefficients, the number of participating clients, the model dimension, and whether client-level or example-level differential privacy is considered.
        \item
        When the secret data is a Gaussian random vector $\N{\bm{0}}{C^2 \bm{I}_K}$, then the normalized minimum \ac{mse} is $S=1-(1+\frac{\sigma^2}{C^2})^{-1}$, giving \ac{mse} security with security level $S$. Again, a larger variance $\sigma^2$ yields stronger security (larger $S$).
    \end{enumerate}
    Nonetheless, the measures are generally not equivalent.
    \ac{dp} is based on the conditional output distribution $p(\bm{z}|\bm{s})$ and measures the indistinguishability of different realizations $\bm{s}^\prime,\bm{s}^{\prime\prime}$ by contrasting their conditional output distributions.
    \ac{mse} security is based on the posterior distribution $p(\bm{s}|\bm{z})$ and measures the posterior's variance. As such, it takes into account the secret's prior distribution implicitly.
\end{remark}

\subsection{Secure Encoder Design}
\label{sec:encoder}
To achieve security against eavesdropping, one common strategy in the literature is to add artificial noise (perturbation signals) prior to signal transmission to hide the true information. Therefore, the encoding scheme can be modified by letting each user $k$ transmit
\begin{equation}
    x_k = \eta h_k^{-1}\gamma_k + w_k,
\end{equation}
where $w_k\in \complexes$ represents the added artificial noise. 
We first discuss the artificial noise design and then return to the power control design.

\begin{remark}
\label{remark:signal_definitions}
    For secure \ac{ota} aggregation design, two slightly different operations exist in the literature: adding noise after scaling (cf. \cite{massny2023secure}) 
    \begin{equation}
    \label{eq:used_noise_scheme}
        x_k^{(A)} = \eta^{(A)} h_k^{-1}\gamma_k + w_k^{(A)},
    \end{equation}
    and adding noise before 
    scaling (cf. \cite{liaoOvertheAirFederatedLearning2022})
    \begin{equation}
        x_k^{(B)} = \eta^{(B)} h_k^{-1}(\gamma_k + w_k^{(B)}).
    \end{equation}

    These representations are equivalent, since one can always find a set of valid parameters (amplitude scaling factor and artificial noise design) in one framework that corresponds to a set of parameters in the other framework such that $x_k^{(A)} = x_k^{(B)}$. 
    In this work, we use the framework in \eqref{eq:used_noise_scheme} since it gives an intuitive interpretation of the system design, i.e., one can fix the power of the real data signal through $\eta$, then use the remaining power $P-(\eta/\abs{h_k})^2$ for security preservation.
\end{remark}

\subsubsection{Correlated Artificial Noise Design}
Adding artificial noise generally increases the approximation error (aggregation noise perceived at the server) since such noise directly enters the estimated computation result. One way to mitigate this effect is to consider spatially correlated (or zero-forced) noise such that the received aggregated artificial noise sums to zero at the server. To encompass a variety of such schemes, we consider a general correlated noise design of the form \begin{equation} 
    \label{eq:noise}
    \vect{w} = (w_1,w_2,\ldots,w_K)^T = \bm{A}\vect{v},
\end{equation}
where $\bm{A}\in \complexes^{K\times m}$ is the noise precoding matrix and 
$\vect{v} \in \complexes^m$ a random vector, for some $m\leq K$.

To facilitate the analysis, we consider Gaussian distributed artificial noise, i.e., $\vect{v} \sim \CN{\bm{0}}{\bm{I}_m}$, as is commonly done in the literature \cite{He2024secure, liaoOvertheAirFederatedLearning2022, OtAFLDP, privacy-for-free, goelGuaranteeingSecrecyUsing2008,khistiGaussianMIMOWiretap2007}. Using this representation, we can describe the artificial noise by its covariance matrix $\bm{Q} = \Cov{\vect{w}} = \bm{A}\bm{A}^H$, where $m$ limits the rank of $\bm{Q}$, or equivalently, the subspace in which $\vect{w}$ lies. 

\begin{remark}
The choice of $\mathbf{A}$ allows us to recover many common artificial noise designs. Setting $\mathbf{A}=\mathbf{I}$ yields uncorrelated noise (cf. \cite{OtAFLDP, privacy-for-free, khistiGaussianMIMOWiretap2007}). A further special case is zero-forced noise, where $\vect{h}^T\vect{w}=0$ (cf. \cite{He2024secure, liaoOvertheAirFederatedLearning2022, goelGuaranteeingSecrecyUsing2008}), which is achieved by choosing $\bm{A}$ such that $\vect{h}^T\bm{A} = \vect{0}$, i.e., by letting the columns of $\bm{A}$ be orthogonal to $\vect{h}\conj$. This requires $m < K$.
\end{remark} 

In practice, users can jointly generate their artificial noise through the following process: the noise matrix $\bm{A}$ is generated at the central server and shared with all users, along with a common random seed that is used to locally generate the same pseudo-random noise vector $\vect{v}$. 
With this, each user $k$ computes their artificial noise as $w_k = \left[ \bm{A}\vect{v}\right]_k$. This process motivates the design choice to make $\bm{A}$ non-square, since smaller $m$ reduces the overhead cost of sharing $\bm{A}$ with all users, at the cost of shrinking its design space.\footnote{
    It might seem preferable to directly share the noise vector $\vect{w}$ with all users since its dimension is smaller than the noise precoding matrix $\bm{A}$. However, in case the source data contain high-dimensional vectors, it would be much more efficient to share $\bm{A}$ instead of $\vect{w}$. Additionally, as will be shown later, $\bm{A}$ can be broadcast openly and thus known by the eavesdroppers while still providing some security guarantees.
    }
    
Nevertheless, for large $K$, broadcasting $\bm{A}$ to all users may still incur non-negligible coordination overhead. While the proposed scheme assumes network-wide correlation, i.e., all $K$ nodes sharing a common $\bm{A}$ and the random seed, more granular correlation structures are also possible. For instance, pairwise or subset-wise noise correlation can be realized by assigning independent seeds and noise precoding matrices to different groups of users.\footnote{With the noise scheme described in (\ref{eq:noise}), this corresponds to $\bm{A}$ having a sparse block structure.} This could potentially be done in a fully decentralized fashion, where subsets of users locally agree on complementary noise terms that cancel at the receiver without explicit coordination from the server. Such hierarchical schemes reduce the coordination overhead associated with distributing $\bm{A}$ and synchronizing the seed, at the expense of fewer degrees of freedom in the joint noise design and, potentially, a degraded privacy guarantee. Characterizing the tradeoff between coordination cost and privacy performance as a function of the correlation granularity, as well as for our proposed scheme with a limitation on $m$, is a natural and interesting direction for future work.
    
\subsubsection{The Effects of Power Control} 
As stated in Remark \ref{remark:signal_definitions}, the amplitude scaling factor $\eta$ controls the division of power between the real data signal and artificial noise, i.e., it controls the trade-off between \ac{mse}-security and \ac{mse}-approximation. Let $\vect{a}_1^T,\ldots,\vect{a}_K^T$ denote the rows of $\bm{A}$, then we have $\EX{\abs{w_k}^2} = \norm{\vect{a}_k}^2$. Consequently, the individual power constraints $\EX{\abs{x_k}^2} \leq P$ yield the following upper bound on the amplitude scaling factor:
\begin{equation}
\label{eq:power_constraint}
    \eta^2 \leq \min_{k\in \{ 1,2,\ldots,K \} } \abs{h_k}^2\left( P - \norm{\vect{a}_k}^2 \right).
\end{equation}
With legitimate users employing the secure encoding design, the server and eavesdroppers receive
\begin{align}
    y &= \sum_{k=1}^K \left(\eta \gamma_k  + h_k w_k\right)  + n_y = \eta s + \vect{h}^T\bm{A}\vect{v} + n_y,\\
    \label{eq:z_l}
    z_\ell &= \eta \sum_{k=1}^K \frac{g_{\ell,k}}{h_k}\gamma_k + \vect{g}_\ell^T\bm{A}\vect{v} + n_{z_\ell} \\
    \nonumber&= \eta s + \eta \sum_{k=1}^K (\frac{g_{\ell,k}}{h_k}-1)\gamma_k + \vect{g}_\ell^T\bm{A}\vect{v} + n_{z_\ell}, \; \;  \forall \ell\in \{1,\ldots,L\} \;,
\end{align}
where $s$ is the intended computation result. Theorem \ref{thm:approximation_level} gives the resulting approximation error at the server.

\begin{theorem}
\label{thm:approximation_level}
Consider the proposed \ac{ota} computation scheme with source data $\bm{\gamma} \sim \CN{\bm{0}}{\bm{I}_K}$. 
The computation result is \ac{mse}-approximate with approximation error 
$$ D =  1 - \frac{\eta^2 K}{\eta^2K + \Vert \vect{h}^T\bm{A}\Vert^2 + \sigma_y^2}.$$

\begin{proof}
See Appendix~\ref{app:th_approximation}.
\end{proof}
\end{theorem}
Based on this result, Corollary \ref{co:eta} gives the achievable design space for the amplitude scaling factor $\eta$, given an upper bound on the approximation error. 
\begin{corollary}
\label{co:eta}
For a given upper bound on the approximation error, $ D\leq \mu$, the amplitude scaling factor must satisfy 
$$\frac{(1-\mu)(\norm{\vect{h}^T\bm{A}}^2 + \sigma_y^2)}{\mu K} \leq \eta^2 \leq \mkern 15mu \min_{\stackrel{\null}{\mathclap{k\in \{ 1,2,\ldots,K \} }}} \mkern 15mu \abs{h_k}^2\left( P - \norm{\vect{a}_k}^2 \right).$$

\end{corollary}

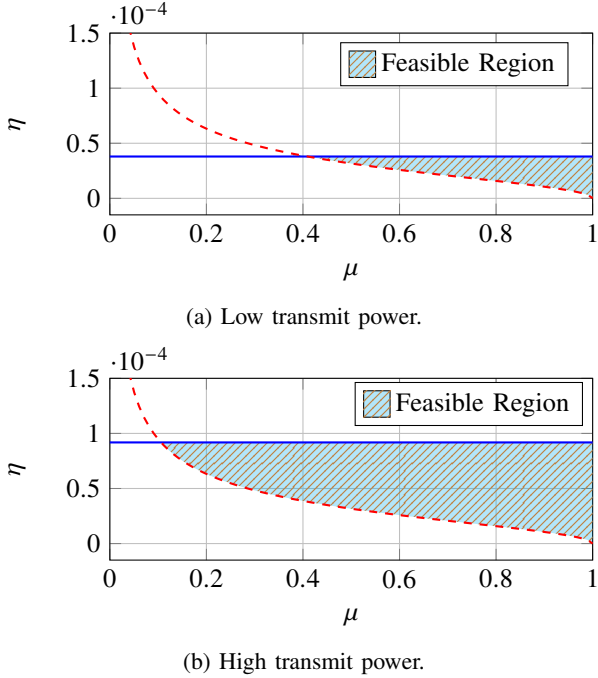
\begin{figure}
    \centering  
    \begin{subfigure}[b]{\linewidth}
        \centering
        \begin{tikzpicture}
            \begin{axis}[
            width=7cm,
            height=4cm,
            xlabel={$\mu$},
            ylabel={$\eta$},
            grid=both,
            xmin=0, xmax=1,
            ymax=1.5e-4,
            yticklabel style={/pgf/number format/fixed},
            legend style={at={(0.95,0.95)}, anchor=north east, legend columns=1},
            ]
            \addlegendimage{
                legend image code/.code={
                    \path[
                        fill=cyan, fill opacity=0.3,
                    ] (-0.5em,-0.5em) rectangle (0.5em,0.5em);
                    \draw[
                        pattern=north east lines,
                        pattern color=brown,
                        draw=none
                    ] (-0.5em,-0.5em) rectangle (0.5em,0.5em);
                }
            }
            \addlegendentry{Feasible Region}
         
            \addplot [color=blue, thick]
            table [x=mu, y=upper, col sep=space] {tikz_data/eta_design_space_P=1.dat};
          
            \addplot [color=red, dashed, thick]
            table [x=mu, y=lower, col sep=space] {tikz_data/eta_design_space_P=1.dat};
            
            \addplot [name path=upper, draw=none]
            table [x=mu, y=upper, col sep=space] {tikz_data/eta_design_space_P=1.dat};
            
            \addplot [name path=lower, draw=none]
            table [x=mu, y=lower, col sep=space] {tikz_data/eta_design_space_P=1.dat};
            
            \addplot[fill=cyan, fill opacity=0.3]
            fill between [
              of=upper and lower,
              soft clip={domain=0.38:1}];
              
            \addplot[pattern=north east lines, pattern color=brown]
            fill between [
              of=upper and lower,
              soft clip={domain=0.38:1}];

            \end{axis}
            
        \end{tikzpicture}
        \caption{Low transmit power ($P=1$).}
    \end{subfigure} 
    \\[0.2cm]
    \begin{subfigure}[b]{\linewidth}
        \centering
        \begin{tikzpicture}
          \begin{axis}[
            width=7cm,
            height=4cm,
            xlabel={$\mu$},
            ylabel={$\eta$},
            grid=both,
            xmin=0, xmax=1,
            ymax=1.5e-4,
            yticklabel style={/pgf/number format/fixed},
          ]

          \addlegendimage{
                legend image code/.code={
                    \path[
                        fill=cyan, fill opacity=0.3,
                    ] (-0.5em,-0.5em) rectangle (0.5em,0.5em);
                    \draw[
                        pattern=north east lines,
                        pattern color=brown,
                        draw=none
                    ] (-0.5em,-0.5em) rectangle (0.5em,0.5em);
                }
            }
            \addlegendentry{Feasible Region}

          \addplot [color=blue, thick]
            table [x=mu, y=upper, col sep=space] {tikz_data/eta_design_space_P=10.dat};
          
          \addplot [color=red, dashed, thick]
            table [x=mu, y=lower, col sep=space] {tikz_data/eta_design_space_P=10.dat};
    
          \addplot [name path=upper, draw=none]
            table [x=mu, y=upper, col sep=space] {tikz_data/eta_design_space_P=10.dat};
    
          \addplot [name path=lower, draw=none]
            table [x=mu, y=lower, col sep=space] {tikz_data/eta_design_space_P=10.dat};
    
          \addplot[fill=cyan, fill opacity=0.3]
            fill between [
              of=upper and lower,
              soft clip={domain=0.11:1}, 
            ];
            \addplot[pattern=north east lines, pattern color=brown]
            fill between [
              of=upper and lower,
              soft clip={domain=0.11:1}, 
            ];
            \end{axis}
        \end{tikzpicture}
        \caption{High transmit power ($P=10$).}
    \end{subfigure}
    \caption{ The shaded area represents the design space for the amplitude scaling factor $\eta$ under different approximation error constraints. The bounding curves are derived from the upper and lower bounds in Corollary \ref{co:eta}. See Section \ref{sec:results} for details on the system parameters used.}
    \label{fig:eta}
\end{figure}

Based on Corollary~\ref{co:eta}, 
the design space of $\eta$ is illustrated in Fig. \ref{fig:eta}. Each point is achievable for some choice of $\bm{A}$. With zero-forced noise, the lower bound is independent of $\mathbf{A}$, thus yielding the largest feasible region. With other designs, the lower bound increases according to the amount of artificial noise that enters the received signal at the server.
This result sheds some light on the intricate relationship between the approximation error and security level. Smaller $\eta$ increases the power allocated to the artificial noise, thus increasing the security level, but it simultaneously increases the approximation error since less power is allocated to the data. To compensate for that, one can design the artificial noise carefully to reduce $\Vert\vect{h}^T\bm{A}\Vert^2$ (e.g., by zero-forcing), limiting the design space of $\bm{A}$, which might in turn decrease the security level. We will continue this discussion in Section \ref{sec:results}.

\subsection{Joint Decoder Design at Multiple Eavesdroppers}
Based on the received signal at each eavesdropper given in \eqref{eq:z_l}, the $L$ eavesdroppers can jointly process their received signals and obtain an estimate of the target function value, denoted by $\hat{s}$.
The security level in Definition \ref{def:security} is then the normalized expected \ac{mse} after joint processing (receive combining). For a given channel realization, the \ac{mmse} estimator 
$$d_{z,{\text{opt}}} = \argmin_{d_z: \complexes^L \to \complexes} \EX{\vert d_z(z_1,\ldots,z_L ; \bm{\Phi}) - s \vert ^2},$$ 
is linear since the observations $\{z_\ell\}_{\ell=1,\ldots, L}$ and the estimation target $s$ are jointly Gaussian \cite[Prop. 3.9]{hajek2015random}.
Therefore, the target function estimate is $\hat{s} = d_z(z_1,\ldots,z_L ; \bm{\Phi}) = \vect{p}^H\vect{z}$, where $\vect{z}=(z_1,\ldots,z_L)^T$ and $\vect{p}=(p_1,\ldots,p_L)^T \in \complexes^L$ is the combining vector. 
Being a linear combination of each eavesdropper's received signal, the estimate can be viewed as the decoded signal at a single \textit{virtual eavesdropper}, which is connected to $L$ spatially distributed eavesdroppers. 

Introducing notations $\vect{n}_z = (n_{z_1},\ldots,n_{z_L})^T$ and 
$$\bm{D}_{\vect{h}} =
  \begin{bmatrix}
    h_{1} & & 0\\
    & \ddots & \\
    0& & h_{K}
  \end{bmatrix} , \hspace{0.2cm} \bm{G} =\begin{bmatrix}
    \text{---} \hspace{-0.2cm} & \vect{g}_1^T & \hspace{-0.2cm} \text{---} \\
    \text{---} \hspace{-0.2cm} & \vect{g}_2^T & \hspace{-0.2cm} \text{---}  \\ & \vdots &\\
    \text{---} \hspace{-0.2cm} & \vect{g}_L^T & \hspace{-0.2cm} \text{---}
\end{bmatrix},$$
we can write the received signal vector $\vect{z}$ at the virtual eavesdropper  as
\begin{equation}
\label{eq:z_vec}
    \vect{z} = \eta \bm{G} \bm{D}_{\vect{h}}^{-1}\bm{\gamma} + \bm{G}\bm{A}\vect{v} + \vect{n}_z. 
\end{equation}
The \ac{mmse} estimator is
\begin{equation}
    \hat{s}_{\text{opt}} = d_{z,{\text{opt}}}(\vect{z}) = \vect{p}_{\text{opt}}^H \vect{z},
\end{equation}
where 
\begin{equation}
\label{eq:popt}
    \vect{p}_{\text{opt}} = \argmin_{ \vect{p}\in \complexes^L} \EX{ \vert \vect{p}^H\vect{z} - s \vert ^2}.
\end{equation}
With this, the security level in Definition \ref{def:security} can be equivalently  written as
\begin{equation}
    S = \frac{\EX{ \vert \vect{p}_{\text{opt}}^H\vect{z} - s \vert ^2}}{\Var{s}}
    = \frac{\min_{ \vect{p}\in \complexes^L} \EX{ \vert \vect{p}^H\vect{z} - s\vert ^2}}{\Var{s}}.
\end{equation}
Finding the optimal eavesdropper strategy, thus, reduces to finding the optimal combining vector, which is presented in the following section.

\subsection{Extension to Multi-Antenna Systems}

Our system model focuses on the case with single-antenna senders and receivers. Here, we briefly discuss extensions to multi-antenna settings.

\subsubsection{Multi-Antenna Receivers}

Suppose that the server is equipped with $N_r > 1$ antennas. The received signal can then be written as
\begin{equation}
\mathbf{y} = \mathbf{H}\mathbf{x} + \mathbf{n}_y,
\end{equation}
where $\mathbf{H} \in \mathbb{C}^{N_r \times K}$ is the channel matrix between the $K$ users and the server, and $\mathbf{n}_y \sim \mathcal{CN}(\mathbf{0}, \sigma_y^2 \mathbf{I})$. The server applies a linear receive combiner $\mathbf{r} \in \mathbb{C}^{N_r}$ to obtain
\begin{equation}
\hat{s} = \mathbf{r}^H \mathbf{y}.
\end{equation}

The additional spatial degrees of freedom provided by multiple antennas can be exploited to improve the estimation performance through appropriate beamforming design. 
However, the artificial noise design must be adapted accordingly. Extending the zero-forcing principle in a per-antenna manner (i.e., enforcing $\mathbf{H}\mathbf{w} = \mathbf{0}$) would impose significant constraints on the noise design. A more effective approach is to jointly design the artificial noise vector $\mathbf{w}$ and the receive combiner $\mathbf{r}$ such that the artificial noise is suppressed after receive combining, i.e., $\mathbf{r}^H \mathbf{H}\mathbf{w} = 0$.

\subsubsection{Multi-Antenna Senders}

Consider now the scenario in which each user is equipped with multiple antennas. Let user $k$ have $N_{t,k} > 1$ antennas, transmitting $\mathbf{x}_k \in \mathbb{C}^{N{t,k}}$. The overall received signal at the server becomes
\begin{equation}
\mathbf{y} = \sum_{k=1}^K \mathbf{H}_k \mathbf{x}_k + \mathbf{n}_y,
\end{equation}
where $\mathbf{H}_k \in \mathbb{C}^{N_r \times N{t,k}}$ is the channel matrix from user $k$ to the server.

The additional transmit spatial degrees of freedom can be exploited to enhance both the security level and aggregation accuracy. In particular, users can apply linear precoders $\vect{v}_k \in \complexes^{N{t,k}}$ to their data such that
\begin{equation}
\mathbf{x}_k = \vect{v}_k \gamma_k + \mathbf{w}_k,
\end{equation}
where $\mathbf{w}_k \in \complexes^{N{t,k}}$ is the $k$-th artificial noise vector.

A key design challenge is to coordinate the artificial noise across multiple users. Fully network-wide coordination to null the collective artificial noise at the server incurs high overhead costs. When each user has multiple transmit antennas ($N_{t,k}$), it becomes feasible for some of the users to independently zero-force their own artificial noise along their effective channel to the server, without requiring coordination with other users. This per-user zero-forcing reduces overhead costs, but the number of antennas directly impacts the spatial degrees of freedom available for artificial noise shaping.

While optimal transmit and receive beamforming strategies have been extensively studied in the \ac{ota} multiple-input multiple-output (MIMO) literature from the perspective of the server's performance\cite{otaZhu, zhu2018mimo, asaad2024joint, wen2019reduced, jing2023transceiver,chen2024over}, the interplay between beamforming design and artificial noise generation, and its subsequent impact on the achieved security level, remains largely unexplored and represents a promising direction for future work.

\subsubsection{Multi-Antenna Eavesdroppers}

Finally, we discuss the case where the eavesdroppers are equipped with multiple antennas. 
By stacking all received signals, the eavesdroppers can again be viewed as a virtual multi-antenna receiver with an increased number of spatial dimensions.

As in the single-antenna case, cooperation among eavesdroppers enables joint post-processing, effectively increasing their spatial degrees of freedom. The presence of multiple antennas at each eavesdropper further amplifies this effect. In particular, the enlarged receive space increases the likelihood that certain spatial directions experience limited artificial noise power, allowing the eavesdroppers to obtain a more accurate estimate of the aggregated signal.

From the perspective of artificial noise design, this extension makes the problem   more challenging. Intuitively, the artificial noise should be structured such that it is suppressed at the server, while remaining effective in degrading the eavesdroppers' observations. However, with an increased number of receive dimensions at the eavesdroppers, it becomes more difficult to ensure that the artificial noise significantly impacts all directions of their effective signal subspace.

\section{Security Level Analysis and Optimal Eavesdropper Strategy}
\label{sec:security_level}
In this section, we evaluate the security level under different side information assumptions and a special case with non-cooperating eavesdroppers (i.e., they are unaware of each other's existence). Additionally, we provide the optimal (eavesdropper) combining vectors when applicable.
The eavesdroppers are assumed to know $\vect{h}$ and $\bm{A}$ but not $\vect{v}$, since $\bm{A}$ is broadcast to all users, whereas $\vect{v}$ is locally generated from a shared source of randomness that is securely available only to legitimate users.

\subsection{Scenario 1: Exact CSI Known}
\label{sec:security_main}
Assuming that the channel coefficients between legitimate users and eavesdroppers are perfectly known to the eavesdroppers ($\bm{\Phi} = \left\{ \vect{g}_1, \ldots, \vect{g}_L\right\}$),
the security level in Definition \ref{def:security} can be computed as follows.

\begin{theorem}
\label{thm:security_level}
Consider the proposed \ac{ota} computation scheme with source data $\bm{\gamma} \sim \CN{\bm{0}}{\bm{I}_K}$ and the cooperative eavesdropper strategy.
The computation result is \ac{mse}-secure with security level 
$$ S = 1 - \frac{1}{K} \vect{m}^H \bm{B}^{-1}\vect{m}, $$
and the optimal combining vector is 
$$\vect{p}_{\text{opt}} = \bm{B}^{-1}\vect{m},$$
where 
\begin{align*} 
\bm{B}&=\bm{G}\bm{A} \bm{A}^H\bm{G}^H + \eta^2 \bm{G} \bm{D}_{\vect{h}}^{-1}\bm{D}_{\vect{h}}^{-H}\bm{G}^H  + \sigma_z^2 \bm{I}_L, \\
\vect{m} &= \eta \bm{G} \bm{D}_{\vect{h}}^{-1} \vect{u} = \begin{bmatrix}
\eta\sum_{k=1}^K g_{1,k}/h_k \\
\eta\sum_{k=1}^K g_{2,k}/h_k \\
\vdots \\
\eta\sum_{k=1}^K g_{L,k}/h_k
 \end{bmatrix}, \hspace{0.2cm} \vect{u} = 
\begin{bmatrix}
1 \\
\vdots \\
1
 \end{bmatrix}.
\end{align*}
\begin{proof}
See Appendix \ref{app:th_security}. 
\end{proof}
\end{theorem}
The matrix $\bm{B}$ consists of three terms, corresponding to the alignment between the artificial noise and the eavesdropper channels, the channel mismatch between legitimate users and eavesdroppers, and the additive noise, respectively. Intuitively, beamforming the artificial noise towards the eavesdropper channels improves the security. 
\begin{remark}
     After optimal combining, the virtual eavesdropper obtains the estimate
\begin{equation}
    \hat{s} = \vect{p}_{\text{opt}}^H \vect{z} = \vect{p}_{\text{opt}}^H \bm{G} \vect{x}  + \vect{p}_{\text{opt}}^H \vect{n}_z = \tilde{\vect{g}}^T \vect{x} + \tilde{n}_z,
\end{equation}
where $\tilde{\vect{g}} = \vect{p}_{\text{opt}}^H\bm{G}$ can be viewed as the effective channel vector and $\tilde{n}_z \sim \CN{0}{\sigma_z^2 \norm{\vect{p}_{\text{opt}}}^2}$ as additive noise. Then, 
using \cite[Th.1]{massny2023secure}, we can equivalently write the security level in Theorem \ref{thm:security_level} as
\begin{equation}
    S = 1 - \dfrac{\frac{\eta^2}{K}\left\vert \sum_{k=1}^K  \dfrac{\tilde{g}_k}{h_k} \right\vert^2}{\eta^2\sum_{k=1}^K  \left\vert \dfrac{\tilde{g}_k}{h_k} \right\vert^2 + \norm{\tilde{\vect{g}}^T \bm{A}}^2 + \sigma_z^2 \norm{\vect{p}_{\text{opt}}}^2}.
\end{equation}
\end{remark}
\subsection{Scenario 2: No Phase Information}
\label{sec:nophase}
It might be unrealistic to assume that the eavesdroppers have perfect knowledge of the channels between them and the legitimate users. Therefore, we now consider the scenario in which only statistical CSI is available. Consequently, the expectations in Definition \ref{def:security} are taken over the randomness in $\bm{G}$ and $\vect{h}$, in addition to other sources of randomness in the system.
We consider random complex fading coefficients $\vect{g}_\ell$, with uniform phase distribution.
In this case, we show that the security level is $S=1$.

\begin{theorem}
\label{thm:statistical}
    If the channel coefficients $g_{\ell,k}=r_{\ell,k} \exp(-j \phi_{\ell,k})$ have a uniformly distributed phase, $\phi_{\ell,k} \sim \mathcal{U}([0,2\pi])$, then the security level is $S=1$.
    
    \begin{proof}
    Note that, for any $\psi$, the distributions of $\bm{G}$ and $e^{j\psi}\bm{G}$ as well as $\bm{\gamma}$ and $e^{-j\psi}\bm{\gamma}$ are the same. Let $p(\cdot)$ denote the probability density function of a continuous random variable. Then, for $\bm{\gamma}^\prime = e^{-j\psi}\bm{\gamma}$, we have $p(\bm{\gamma}) = p(\bm{\gamma}^\prime)$ and $p(\vect{z} \vert \bm{\gamma}) = p(\vect{z} \vert \bm{\gamma}^\prime)$, and therefore $p(\bm{\gamma}^\prime|\vect{z}) = p(\vect{z}|\bm{\gamma}^\prime)p(\bm{\gamma}^\prime)/p(\vect{z}) = p(\vect{z}|\bm{\gamma})p(\bm{\gamma})/p(\vect{z})=p(\bm{\gamma} |\vect{z})$. Consequently, the \ac{mmse} estimator of the data is $\EX{\bm{\gamma}|\vect{z}} = \EX{\bm{\gamma}^\prime|\vect{z}} = e^{-j\psi} \EX{\bm{\gamma}|\vect{z}}$ for arbitrary $\psi$. The only possibility is $\EX{\bm{\gamma}|\vect{z}}=\bm{0}$, from which it follows that the \ac{mmse} estimator of $s$ is $\EX{s | \vect{z}} = \boldsymbol{1}^T \EX{\boldsymbol{\gamma} | \vect{z}} = 0$. The security level is thus $S= \frac{1}{\Var{s}}\EX{\vert 0-s\vert^2} = \Var{s}/\Var{s} = 1$.
    \end{proof}
\end{theorem}
Observing that $1$ is the maximum achievable security level, we conclude that when only channel statistics are known to eavesdroppers, there is no need to add extra artificial noise to improve \ac{mse}-security in our considered setting. 
Therefore, in Section \ref{sec:optnoise}, we will focus on the case with known CSI when deriving the optimal artificial noise design.

\subsection{Non-cooperative Eavesdroppers}
\label{sec:security_noncoop}
In the event that the eavesdroppers are not cooperating, we define the security level as
\begin{equation}
    S=\frac{1}{\Var{s}} \min_{\ell\in (1,\ldots,L)} \left( \min_{d_z: \complexes \to \complexes} \EX{ \vert d_z(z_\ell;\Phi) - s \vert ^2 \vert \bm{\Phi}} \right).
\end{equation}
That is, the security at the eavesdropper with the lowest \ac{mse}. Under the same side information assumptions as in Section \ref{sec:security_main}, using \cite[Th.1]{massny2023secure}, it can be computed as 
\begin{equation}
\label{eq:security_noncoop}
    S =  \min_{ \ell \in (1,\ldots,L)} \left( 1 - \dfrac{ \frac{\eta^2}{K}\left\vert \displaystyle\sum_{k=1}^K  \dfrac{g_{k,\ell}}{h_k} \right\vert^2}{\eta^2 \displaystyle\sum_{k=1}^K  \left\vert \dfrac{g_{k,\ell}}{h_k} \right\vert^2 + \norm{\vect{g}_\ell^T \bm{A}}^2 + \sigma_{z_\ell}^2} \right) . 
\end{equation}

\subsection{Inherent Security and Impact of Eavesdropper Cooperation}

\begin{figure}
    \centering  
    \begin{subfigure}[b]{\linewidth}
        \centering
        \begin{tikzpicture}
            \begin{axis}[
                xlabel={Number of Eavesdroppers ($L$)},
                ylabel={Normalized MSE},
                legend style={
                  at={(0.5,1.05)},
                  anchor=south,
                  cells={align=left},
                  legend columns=3
                },
                mark options={solid},
                grid=both,
                width=7cm,
                height=5cm,
                xmin=1, xmax=15,
                ymin=0, ymax=1, 
            ]
              \addplot[
                color=blue,
                dashed,
                mark=*,
                mark size=2pt,
              ] 
              table [x=L, y=D, col sep=space] {tikz_data/complex_channel.dat};
              \addlegendentry{$D$}
              
              \addplot[
                color=red,
                dotted,
                mark=square*,
                thick
              ] 
              table [x=L, y=S1, col sep=space] {tikz_data/complex_channel.dat};
              \addlegendentry{$S_\text{coop}$}
        
              \addplot[
                color=black,
                solid,
                mark=triangle*,
                mark size=3pt,
                thick
              ] 
              table [x=L, y=S2, col sep=space] {tikz_data/complex_channel.dat};
              \addlegendentry{$S_\text{noncoop}$}
            \end{axis}
          \end{tikzpicture}
          \caption{With complex-valued channels.}
          \label{fig:eavesdropper_strategy_nonoise_complex_channels}
    \end{subfigure} 
    \\[0.2cm]
    \begin{subfigure}[b]{\linewidth}
        \centering
        \begin{tikzpicture}
            \begin{axis}[
                xlabel={Number of Eavesdroppers ($L$)},
                ylabel={Normalized MSE},
                legend style={
                  at={(0.5,1.05)},
                  anchor=south,
                  cells={align=left},
                  legend columns=3
                },
                mark options={solid},
                grid=both,
                width=7cm,
                height=5cm,
                xmin=1, xmax=15, 
                ymin=0, ymax=1, 
            ]
              \addplot[
                color=blue,
                dashed,
                mark=*,
                thick
              ] 
              table [x=L, y=D, col sep=space] {tikz_data/real_channel.dat};
              \addlegendentry{$D$}
              
              \addplot[
                color=red,
                dotted,
                mark=square*,
                thick
              ] 
              table [x=L, y=S1, col sep=space] {tikz_data/real_channel.dat};
              \addlegendentry{$S_\text{coop}$}
        
              \addplot[
                color=black,
                solid,
                mark=triangle*,
                mark size=3pt,
                thick
              ] 
              table [x=L, y=S2, col sep=space] {tikz_data/real_channel.dat};
              \addlegendentry{$S_\text{noncoop}$}
            \end{axis}
          \end{tikzpicture}
          \caption{With real-valued channels.}
          \label{fig:eavesdropper_strategy_nonoise_real_channels}
    \end{subfigure}
    \caption{Estimation accuracy of the server, cooperative eavesdroppers and non-cooperative eavesdroppers, in a scenario with no added artificial noise.}
    \label{fig:eavesdropper_strategy_nonoise}
\end{figure}
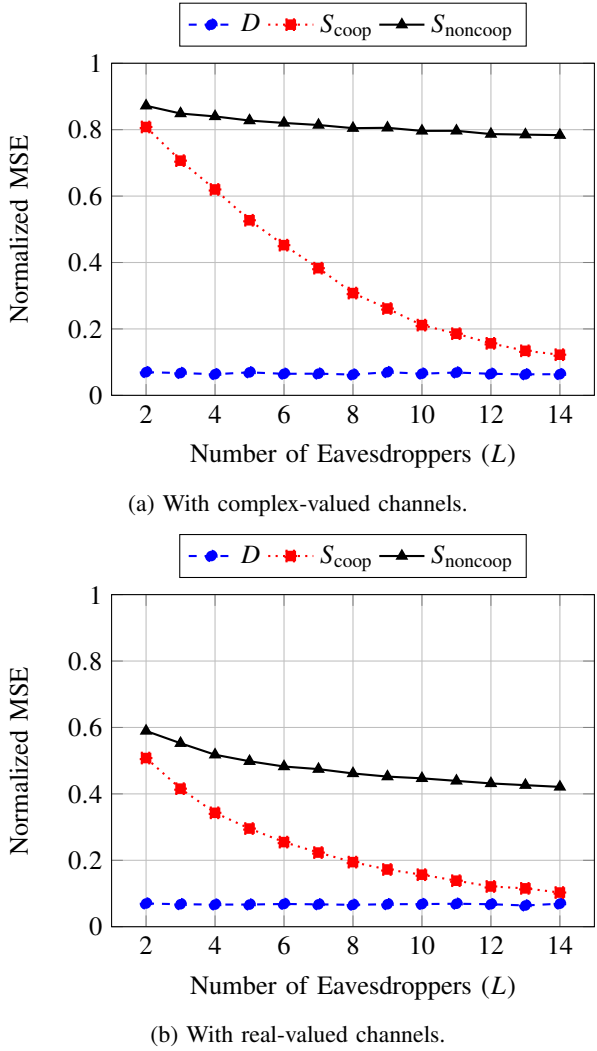

First, we simulate the system without artificial noise ($\bm{A}=0$) to evaluate its inherent security. The system consists of $K=10$ users and a varying number of eavesdroppers. A detailed description of the simulation environment is provided in Section \ref{sec:results}.
Fig. \ref{fig:eavesdropper_strategy_nonoise_complex_channels} shows the approximation error and security levels in the cooperative and non-cooperative cases, computed according to Theorem \ref{thm:security_level} and (\ref{eq:security_noncoop}), respectively.
It can be observed that the security level decreases significantly as the number of cooperating eavesdroppers $L$ increases.
Interestingly, the non-cooperative security level experiences only minimal degradation as $L$ increases. Since the minimum security level across the eavesdroppers remains relatively constant with the increase in $L$, we conclude that all eavesdroppers must exhibit similarly poor aggregation accuracy.

In some earlier studies, such as \cite{ massny2023secure}, the channel coefficients were modeled as real-valued random numbers. In this case, there is no phase misalignment in the received signals, and individual eavesdroppers can attain higher estimation accuracy.
In practice, when I/Q modulation is used, the channel coefficients are modeled as complex-valued random numbers, where the phase of the signal carries information about the data symbols. Misaligned channel phases between the legitimate user channels and eavesdropper channels lead to a larger value in the numerator of (\ref{eq:security_noncoop}). Consequently, the security level obtained with the complex-valued channel model is inherently higher than the real-valued case, as illustrated by the comparison between Fig. \ref{fig:eavesdropper_strategy_nonoise_complex_channels} and Fig. \ref{fig:eavesdropper_strategy_nonoise_real_channels}.  

Based on these results, we conclude that there is significant inherent security in \ac{ota} systems when eavesdroppers are not cooperating, induced by phase misalignment between the legitimate user and eavesdropper channels.
Despite this, cooperation among distributed eavesdroppers evidently allows them to exploit their diversity effectively, posing a significant security threat as $L$ increases.\footnote{Without channel noise, $L=K$ eavesdroppers can perfectly recover the source data values from $K$ legitimate users, achieving error-free estimation.}
This motivates the introduction of carefully designed artificial noise, which aims at sustaining a performance gap between the server and eavesdroppers. The choice of $\eta$ also affects this performance gap. However, in this work we consider that $\eta$ has been chosen to satisfy some approximation error requirement, and focus on optimizing the artificial noise.

\section{Artificial Noise Design}
\label{sec:optnoise}
In this section, we address the question of how the artificial noise should be designed. The problem essentially reduces to designing the noise precoding matrix $\bm{A}\in \complexes^{K \times m}$ for some $m<K$. Despite the inherent security in the non-cooperative eavesdropper case, we provide an artificial noise design that optimizes the security in this scenario and subsequently shows that it achieves good security performance even in the cooperative case. 

The security level in the non-cooperative case given in \eqref{eq:security_noncoop} is maximized by solving

\begin{subequations} 
\label{eq:opt}
\begin{align}
 \max_{\bm{A}} ~~  &  \min_{\ell\in (1,\ldots,L)}
    \dfrac{\eta^2 \displaystyle\sum_{k=1}^K  \left\vert \dfrac{g_{k,\ell}}{h_k} \right\vert^2 + \norm{\vect{g}_\ell^T \bm{A}}^2 + \sigma_{z_\ell}^2}{\left\vert \displaystyle\sum_{k=1}^K  \dfrac{g_{k,\ell}}{h_k} \right\vert^2} \tag{\theequation a} \\
 \textrm{subject to} \hspace{0.2cm} ~~ &\norm{\vect{a}_k}^2 \leq P - \eta^2/\abs{h_k}^2, \forall k \in \{1,2,\ldots,K\}, \tag{\theequation b} \label{eq:opt_power} \\ 
 &  \norm{\vect{h}^T\bm{A}}^2\leq \frac{\mu K \eta^2}{(1-\mu)} - \sigma_y^2, \tag{\theequation c} \label{eq:opt_feasible}
\end{align} 
\end{subequations}
where the constraints in (\ref{eq:opt_power}) and (\ref{eq:opt_feasible}) correspond to the power and approximation error constraints derived from the upper and lower bounds in Corollary \ref{co:eta}, respectively.
At each individual eavesdropper, the security is given as an affine function of the received noise power  $\norm{\vect{g}_\ell^T \bm{A}}^2$. The server maximizes the minimum security level across the eavesdroppers by beamforming the artificial noise in the direction of different eavesdroppers proportionally to their channel misalignments. In addition, there are two bias terms. The first term, $\eta^2\sum_{k=1}^K  \left\vert \dfrac{g_{k,\ell}}{h_k} \right\vert^2 / \left\vert \sum_{k=1}^K  \dfrac{g_{k,\ell}}{h_k} \right\vert^2$, increases when the phase between $\vect{g}_\ell$ and $\vect{h}$ are more misaligned. The second term, $\sigma_{z_\ell}^2 / \left\vert \sum_{k=1}^K  \dfrac{g_{k,\ell}}{h_k} \right\vert^2$, decreases as the eavesdroppers' \ac{snr} increases. 
This interpretation suggests that we should focus the artificial noise towards eavesdroppers whose channel vectors are more aligned with $\vect{h}$ and who have relatively good channel conditions.

The problem in (\ref{eq:opt}) can  be transformed to
\begin{subequations}
\label{eq:opt_with_t}
\begin{align}
 \max_{t,\bm{A}} ~~  &  t \tag{\theequation a}\\
 \textrm{subject to} \nonumber\\ \hspace{-0.2cm} ~~ &\dfrac{\eta^2 \displaystyle\sum_{k=1}^K  \left\vert \dfrac{g_{k,\ell}}{h_k} \right\vert^2 + \norm{\vect{g}_\ell^T \bm{A}}^2 + \sigma_{z_\ell}^2}{\left\vert \displaystyle\sum_{k=1}^K  \dfrac{g_{k,\ell}}{h_k} \right\vert^2} \geq t ,\; \forall \ell \in\{1,\ldots ,L \} \tag{\theequation b}\\ 
 &  \norm{\vect{a}_k}^2 \leq P - \eta^2/\abs{h_k}^2, \forall k \in \{1,2,\ldots,K\}, \tag{\theequation c} \\
 & \norm{\vect{h}^T\bm{A}}^2\leq \frac{\mu K \eta^2}{(1-\mu)} - \sigma_y^2.\tag{\theequation d} \label{eq:opt_with_t_feasible}
\end{align} 
\end{subequations}
Since the constraints are non-convex in $\bm{A}$, we propose a structured noise design that makes (\ref{eq:opt_with_t}) efficiently solvable. Furthermore, the proposed design is zero-forcing, thereby inherently satisfying the approximation error constraint in (\ref{eq:opt_with_t_feasible}), assuming that the approximation error bound $\mu$ is achievable (in the noiseless case).\footnote{The optimization problem can be solved (in $\bm{Q} = \bm{A} \bm{A}^H$) using semidefinite relaxation. However, the low rank constraint on $\bm{Q}$, imposed by the tall shape of $\bm{A}$ ($m < K$), complicates either the problem (if a rank constraint is imposed), or the extraction of $\bm{A}$ from the potentially full rank $\bm{Q}$.
}

\subsection{Proposed Noise Design}
We consider zero-forcing noise designs. Although potentially sub-optimal, we motivate this choice by the observation that zero-forcing noise designs are among the top performers with regard to the security-approximation tradeoff (cf. Section \ref{sec:results}). 
In the spirit of~\cite{massny2023secure},  we choose $\bm{A} \in \mathbb{C}^{K\times (K-1)}$ in the form 
\begin{align}
\label{eq:noise_starting_design}
    \bm{A} &= 
    \begin{bmatrix}
    \begin{array}{c c c}
        \vspace{-0.4cm}
         & \hspace{0.5cm}\bm{I}_{K-1} &  \\
          &  & \\
           \hline
        -\frac{h_1}{h_K} & \hdots & -\frac{h_{K-1}}{h_K}
    \end{array}    
    \end{bmatrix} 
    \begin{bmatrix}
    \begin{array}{c c c}
        \sqrt{\lambda_1} & & 0\\
        & \ddots & \\
        0& & \sqrt{\lambda_{K-1}}
    \end{array}
    \end{bmatrix} \nonumber \\
  &= 
  \begin{bmatrix}
  \begin{array}{c c c}
    \vspace{-0.4cm}
         & \hspace{1cm} \bm{\Lambda}^{1/2} &  \\
          &  & \\
           \hline
        - \sqrt{\lambda_{1}}\frac{h_1}{h_K} & \hdots & -\sqrt{\lambda_{K-1}}\frac{h_{K-1}}{h_K}
    \end{array}
    \end{bmatrix},
\end{align}
where $\bm{\Lambda}^{1/2} = \text{diag}(\sqrt{\lambda_{1}}, \ldots, \sqrt{\lambda_{K-1}})$. 
This noise design satisfies the zero-forcing condition, i.e., $\vect{h}^T\bm{A}=\bm{0}$, while allowing control over the power allocation through $(\lambda_1, \ldots, \lambda_{K-1})$. Note that the responsibility to ensure zero-forcing falls on a single user. Without loss of generality, we assume that user $K$ has the best channel condition and thus the most energy to allocate to the artificial noise. 
One key advantage of this design is that it performs particularly well when there is a large imbalance in the power allocated to artificial noise at different users. One such case is if $\eta$ is chosen close to the upper bound. In this case, the user with the worst channel allocates little power to the artificial noise, forcing all elements of the corresponding row of $\bm{A}$ to be small. However, to preserve its zero-forcing nature, the column span must not change in the process. With the proposed design, one row can be made arbitrarily small without limiting the noise generation at the other users. It may even enable them to transmit additional noise, as the $K$-th user has one less noise signal to compensate for.
Conversely, a non-sparse (but still zero-forcing) design would necessitate scaling down the entire matrix and thus the artificial noise power of all users.

The design also offers advantages in terms of communication overhead. 
The general problem of broadcasting $\bm{A}$ ($K^2 - K$ symbols) reduces to broadcasting $\bm{\Lambda}^{1/2}$ ($K-1$ symbols) and sharing $\vect{h}$ ($K-1$ symbols, excluding $h_K$) with the $K$-th user.

With the proposed design, we have
\begin{align}
 \Vert \vect{g}_\ell^T \bm{A} \Vert^2 &= \sum_{i=1}^{K-1} \lambda_i \abs{g_{\ell,i} - g_{\ell,K}\frac{h_i}{h_K}}^2  \; \forall \ell, \\
 \Vert \vect{a}_k \Vert^2 &= 
 \begin{cases}
\lambda_k \hspace{1.7cm},  \text{ if }   k \neq K\\
\sum_{i=1}^{K-1} \lambda_i \abs{\frac{h_i}{h_K}}^2, \text{ otherwise}
\end{cases}.
\end{align}
Then, it follows that 
\begin{equation}
    \dfrac{\eta^2 \displaystyle\sum_{k=1}^K  \left\vert \dfrac{g_{k,\ell}}{h_k} \right\vert^2 + \norm{\vect{g}_\ell^T \bm{A}}^2 + \sigma_{z_\ell}^2}{\left\vert \displaystyle\sum_{k=1}^K  \dfrac{g_{k,\ell}}{h_k} \right\vert^2} = \alpha_\ell + \sum_{i=1}^{K-1} \beta_{\ell,i}\lambda_i,
\end{equation}
where 
\begin{equation}
    \alpha_\ell = \frac{\eta^2 \displaystyle\sum_{k=1}^K  \left\vert \dfrac{g_{k,\ell}}{h_k} \right\vert^2 + 
     \sigma_z^2}{\left\vert \displaystyle\sum_{k=1}^K  \dfrac{g_{k,\ell}}{h_k} \right\vert^2} \;, \hspace{0.2cm} \beta_{\ell,i} = \frac{\abs{g_{\ell,i} - g_{\ell,K}\frac{h_i}{h_K}}^2}{\left\vert \displaystyle\sum_{k=1}^K  \dfrac{g_{k,\ell}}{h_k} \right\vert^2}.
 \end{equation}
The original max-min problem can be transformed into the following optimization problem:
\begin{subequations} \label{eq:proposed}
\begin{align}
 \max_{t,\lambda_1 ,\ldots, \lambda_{K-1}} ~~  &  t \tag{\theequation a} \label{opt:objective} \\
 \textrm{subject to} \hspace{0.2cm} ~~ &\alpha_\ell +\sum_{i=1}^{K-1} \beta_{\ell,i}\lambda_i \geq t ,\; \forall \ell \in\{1,\ldots ,L \} \tag{\theequation b} \label{opt:terms}\\ 
 &  \lambda_k \leq P - \eta^2/\abs{h_k}^2  ,\; \forall k\in\{1,\ldots,K-1\} \tag{\theequation c} \label{opt:power_1}\\
 & \sum_{i=1}^{K-1} \lambda_i \abs{\frac{h_i}{h_K}}^2 \leq P - \eta^2 / \abs{h_K}^2 \tag{\theequation d} \label{opt:power_2}\\
 & \lambda_i \geq 0 , \; \forall i\in\{1,\ldots,K-1\}. \tag{\theequation e}  
\end{align} 
\end{subequations}
The constraints in \eqref{opt:power_1} and \eqref{opt:power_2} correspond to the power constraints. Given that both the objective and the constraints are linear in $\lambda_1, \dots, \lambda_{K-1}$, the problem can be efficiently solved using linear programming techniques.

\subsection{Distributing the Zero-forcing Responsibility}
Additionally, we consider an extension where, rather than relying on a single user to ensure zero-forcing, this responsibility is shared among a subset of users $\mathcal{Z}$ with cardinality $N=|\mathcal{Z}|$. 
Without loss of generality, we assume that the users are ordered such that the first $K-N$ users generate independent noise, and the last $N$ users share the zero-forcing responsibility.
Then, we construct $\bm{A} = \bm{A}^\prime \bm{\Lambda}^{1/2} \in \complexes^{K \times (K- N)}$, where $\bm{\Lambda}^{1/2} = \text{diag}(\sqrt{\lambda_{1}}, \ldots, \sqrt{\lambda_{K-N}})$ and
\begin{equation}
    \bm{A}^\prime = 
    \begin{bmatrix}
    \begin{array}{c c c}
        \vspace{-0.4cm}
        & \bm{I}_{K-N} &  \\
        &  & \\
         \hline \\
         -\frac{h_1}{h_{K-N+1}}d_{K-N+1} & \hdots & -\frac{h_{K-N}}{h_{K-N+1}} d_{K-N+1} \\
        \vdots & \vdots & \vdots \\
         -\frac{h_1}{h_K}d_K & \hdots & -\frac{h_{K-N}}{h_K}d_K
    \end{array}
    \end{bmatrix}.
\end{equation}
The weights $d_{K-N+1},\ldots,d_K$ control each user's contribution to ensuring zero-forcing and must satisfy $\sum_{k=K-N+1}^K d_k = 1$. One option is to choose the weights proportional to the available power, i.e., 
\begin{equation}
    d_k = \frac{P-\eta^2/\abs{h_k}^2}
        {NP -\eta^2 \! \! \! \! \! \! \displaystyle\sum_{j=K-N+1}^K \! \! \! \! \! \!\abs{h_j}^{-2}  }.
\end{equation}
Following the same methodology as in the $N=1$ case, we optimize $\bm{A}$ by solving
\begin{subequations} \label{eq:extension}
\begin{align}
 \max_{\lambda_1 ,\ldots, \lambda_{K-N}} ~~  &  t \tag{\theequation a} \\
 \textrm{subject to} \hspace{0.2cm} ~~ &\alpha_\ell +\sum_{i=1}^{K-N} \beta_{\ell,i}^\prime \lambda_i \geq t ,\; \forall \ell \in\{1\hdots,L \} \tag{\theequation b} \\ 
 &  \lambda_k \leq P - \eta^2/\abs{h_k}^2  ,\; \forall k\in\{1,\ldots,K\} \setminus \mathcal{Z} \tag{\theequation c} \\
 & \sum_{i=1}^{K-N} \lambda_i \abs{d_k\frac{h_i}{h_k}}^2 \leq P - \eta^2 / \abs{h_K}^2, \forall k\in \mathcal{Z} \tag{\theequation d} \\
 & \lambda_i \geq 0 , \; \forall i\in\{1,\ldots,K-N\}, \tag{\theequation e}  
\end{align}
\end{subequations}
where
\begin{equation}
 \beta_{\ell,i}^\prime = \frac{\abs{g_{\ell,i} - \displaystyle\sum_{j=K-N +1}^K \!\!\!g_{\ell,j}\frac{h_i}{h_j} d_j}^2}{\left\vert \displaystyle\sum_{k=1}^K  \dfrac{g_{k,\ell}}{h_k} \right\vert^2}.
 \end{equation}
 This design allows for a larger magnitude of artificial noise, at the cost of restricting the subspace in which the noise is generated, by reducing the rank of $\bm{A}$. In Section \ref{sec:results}, we illustrate the combined impact of these two aspects on the security level. Joint optimization of user selection ($\mathcal{Z}$), zero-forcing weights ($d_{K-N+1},\ldots,d_K$) and power allocation ($\bm{\Lambda}^{1/2}$) is left for future work.

\section{Simulation Results}
\label{sec:results}
In this section, we evaluate the performance of the proposed artificial noise design for secure \ac{ota} computation, examine the effects of power control, and discuss the tradeoff between accuracy and security.  
The simulated system consists of $K=10$ users and $L=5$ eavesdroppers uniformly and randomly distributed in a disk of radius $100\,$m. The server is located in the center of the disk, and neither legitimate users nor eavesdroppers are located within $1\,$m of the server or of each other. All results are averaged over at least $100$ realizations of random positions. Channel coefficients are modeled as random variables following $\CN{0}{d^{-4}}$, where $d$ is the distance of the communication link. Stated \ac{snr} values are the average \ac{snr} with $100\,$m link distance. 
When unspecified, system parameters are configured to achieve an \ac{snr} of $10 \, \text{dB}$.
We consider the scenario where all small-scale fading coefficients are larger than $0.1$, which can be achieved by only scheduling users with sufficiently good channel conditions to participate in the aggregation.
\begin{figure}[!t]
    \centering  
    \begin{subfigure}[b]{\linewidth}
        \centering
        \begin{tikzpicture}[spy using outlines={rectangle, magnification=5, height=1.5cm, width=2.5cm, connect spies}]

          \begin{axis}[
            xlabel={SNR [dB]},
            ylabel={Normalized MSE},
            xticklabel={$\pgfmathparse{10*\tick}\pgfmathprintnumber{\pgfmathresult}$},
            xmode=log,
            log basis x=10,
            xminorticks=false,
            grid=major,
            width=8cm,
            height=6cm,
            every axis plot/.append style={line width=0.8pt},
            xmin=0.006,
            xmax=175,
            legend style={
              at={(0.5,1.05)},
              anchor=south,
              cells={anchor=west},
              legend columns=2,
              column sep=0.1cm,
            },
            mark options={solid},
          ]
            \addplot[black, dashed, mark=*] table [x=P, y=S2_none, col sep=space] {tikz_data/noise_comp.dat};
            \addlegendentry{$S_{\text{noncoop}}$: none}
            
            \addplot[black, thick, mark=*] table [x=P, y=D_none, col sep=space] {tikz_data/noise_comp.dat};
            \addlegendentry{$D$: none}
        
            \addplot[red, dashed, mark=triangle] table [x=P, y=S2_uc, col sep=space] {tikz_data/noise_comp.dat};
            \addlegendentry{$S_{\text{noncoop}}$: signal-level}
            
            \addplot[red, thick, mark=triangle] table [x=P, y=D_uc, col sep=space] {tikz_data/noise_comp.dat};
            \addlegendentry{$D$: signal-level}

            \addplot[orange, dashed, mark=o] table [x=P, y=S2_datalevel, col sep=space] {tikz_data/noise_comp_extra.dat};
            \addlegendentry{$S_{\text{noncoop}}$: data-level}
            
            \addplot[orange, thick, mark=o] table [x=P, y=D_datalevel, col sep=space] {tikz_data/noise_comp_extra.dat};
            \addlegendentry{$D$: data-level}
        
            \addplot[green, dashed, mark=square] table [x=P, y=S2_zfr, col sep=space] {tikz_data/noise_comp.dat};
            \addlegendentry{$S_{\text{noncoop}}$: zeroforcing}
        
            \addplot[green, thick, mark=square] table [x=P, y=D_zfr, col sep=space] {tikz_data/noise_comp.dat};
            \addlegendentry{$D$: zeroforcing}
        
            \addplot[cyan, dashed, mark=x] table [x=P, y=S2_zfo, col sep=space] {tikz_data/noise_comp.dat};
            \addlegendentry{$S_{\text{noncoop}}$: proposed}
        
            \addplot[cyan, thick, mark=x] table [x=P, y=D_zfo, col sep=space] {tikz_data/noise_comp.dat};
            \addlegendentry{$D$: proposed}
            
            \coordinate (spy point) at (axis cs:10,0.93);
            \coordinate (spy view) at (axis cs:0.08,0.19); 
            \spy on (spy point) in node [fill=white] at (spy view);
        
          \end{axis}
        \end{tikzpicture}
        \caption{With non-cooperative eavesdroppers.}
        \label{fig:Res2_noncoop}
    \end{subfigure}
    \\[0.2cm]
    \begin{subfigure}[b]{\linewidth}
        \centering
        \begin{tikzpicture}[spy using outlines={rectangle, magnification=5, height=1.5cm, width=2.5cm, connect spies}]

          \begin{axis}[
            xlabel={SNR [dB]},
            ylabel={Normalized MSE},
            xticklabel={$\pgfmathparse{10*\tick}\pgfmathprintnumber{\pgfmathresult}$},
            xmode=log,
            log basis x=10,
            xminorticks=false,
            grid=major,
            width=8cm,
            height=6cm,
            every axis plot/.append style={line width=0.8pt},
            xmin=0.006,
            xmax=175,
            legend style={
              at={(0.5,1.05)},
              anchor=south,
              cells={anchor=west},
              legend columns=2,
              column sep=0.1cm,
            },
            mark options={solid},
          ]
        
            \addplot[black, dashed, mark=*] table [x=P, y=S1_none, col sep=space] {tikz_data/noise_comp.dat};
            \addlegendentry{$S_{\text{coop}}$: none}
        
            \addplot[black, thick, mark=*] table [x=P, y=D_none, col sep=space] {tikz_data/noise_comp.dat};
            \addlegendentry{$D$: none}
        
            \addplot[red, dashed, mark=triangle] table [x=P, y=S1_uc, col sep=space] {tikz_data/noise_comp.dat};
            \addlegendentry{$S_{\text{coop}}$: signal-level}
        
            \addplot[red, thick, mark=triangle] table [x=P, y=D_uc, col sep=space] {tikz_data/noise_comp.dat};
            \addlegendentry{$D$: signal-level}

            \addplot[orange, dashed, mark=o] table [x=P, y=S1_datalevel, col sep=space] {tikz_data/noise_comp_extra.dat};
            \addlegendentry{$S_{\text{coop}}$: data-level}
            
            \addplot[orange, thick, mark=o] table [x=P, y=D_datalevel, col sep=space] {tikz_data/noise_comp_extra.dat};
            \addlegendentry{$D$: data-level}
            
            \addplot[green, dashed, mark=square] table [x=P, y=S1_zfr, col sep=space] {tikz_data/noise_comp.dat};
            \addlegendentry{$S_{\text{coop}}$: zeroforcing}
        
            \addplot[green, thick, mark=square] table [x=P, y=D_zfr, col sep=space] {tikz_data/noise_comp.dat};
            \addlegendentry{$D$: zeroforcing}
            
            \addplot[cyan, dashed, mark=x] table [x=P, y=S1_zfo, col sep=space] {tikz_data/noise_comp.dat};
            \addlegendentry{$S_{\text{coop}}$: proposed}
        
            \addplot[cyan, thick, mark=x] table [x=P, y=D_zfo, col sep=space] {tikz_data/noise_comp.dat};
            \addlegendentry{$D$: proposed}
            
            \coordinate (spy point) at (axis cs:10,0.88); 
            \coordinate (spy view) at (axis cs:0.08,0.19); 
            \spy on (spy point) in node [fill=white] at (spy view);
          \end{axis}
        \end{tikzpicture}
        \caption{With cooperative eavesdroppers.}
        \label{fig:Res2_coop}
    \end{subfigure} 
    
    \caption{Performance comparison between the server and eavesdroppers, for different noise designs.}
    \label{fig:Res2}
\end{figure}
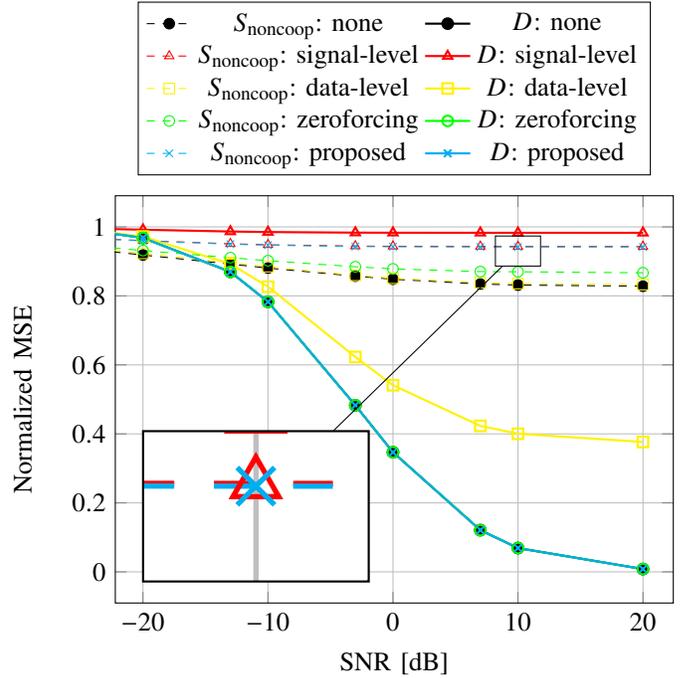
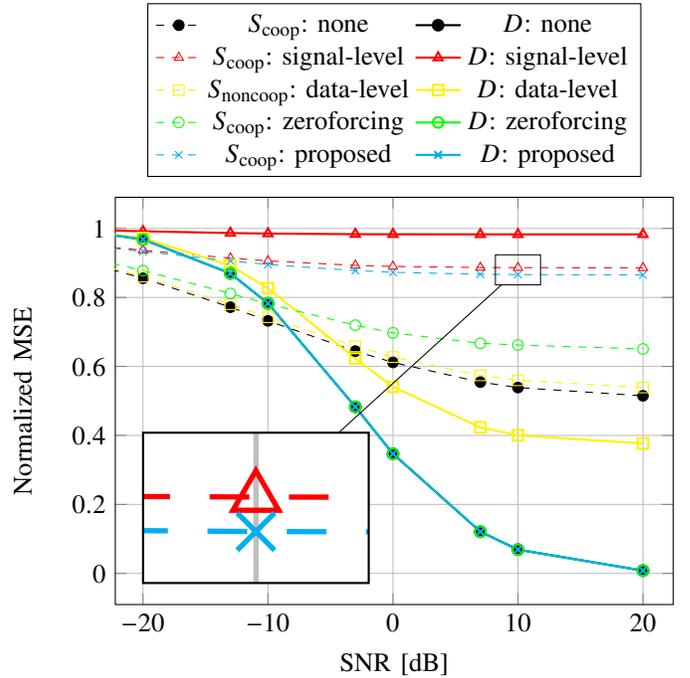

\begin{figure}[!t]
\centering
\begin{tikzpicture}[]

  \begin{axis}[
    xlabel={SNR [dB]},
    ylabel={$S_\text{noncoop}- S_\text{coop}$},
    xticklabel={$\pgfmathparse{10*\tick}\pgfmathprintnumber{\pgfmathresult}$},
    xmode=log,
    log basis x=10,
    xminorticks=false,
    grid=major,
    ymin=0,
    width=7cm,
    height=5cm,
    every axis plot/.append style={line width=0.8pt},
    xmin=0.006,
    xmax=175,
    legend style={
      at={(0.5,1.05)},
      anchor=south,
      cells={align=left},
      legend columns=2
    },
    mark options={solid},
  ]

    \addplot[black, solid, mark=*] table [x=P, y expr=\thisrow{S2_none} - \thisrow{S1_none}, col sep=space] {tikz_data/noise_comp.dat};
    \addlegendentry{none}

    \addplot[red, solid, mark=triangle*, mark size=3pt] table [x=P, y expr=\thisrow{S2_uc} - \thisrow{S1_uc}, col sep=space] {tikz_data/noise_comp.dat};
    \addlegendentry{signal-level}

    \addplot[orange, solid, mark=square*] table [x=P, y expr=\thisrow{S2_datalevel} - \thisrow{S1_datalevel}, col sep=space] {tikz_data/noise_comp_extra.dat};
    \addlegendentry{data-level}

    \addplot[green, solid, mark=o] table [x=P, y expr=\thisrow{S2_zfr}-\thisrow{S1_zfr}, col sep=space] {tikz_data/noise_comp.dat};
    \addlegendentry{zeroforcing}

    \addplot[cyan, solid, mark=x, mark size=3pt] table [x=P, y expr=\thisrow{S2_zfo}-\thisrow{S1_zfo}, col sep=space] {tikz_data/noise_comp.dat};
    \addlegendentry{proposed}

  \end{axis}
\end{tikzpicture}
\caption{Security gap between cooperative and non-cooperative eavesdroppers. Each curve is the difference between the non-cooperative and cooperative security level for that given noise scheme.}
\label{fig:Res2_gap}
\end{figure}
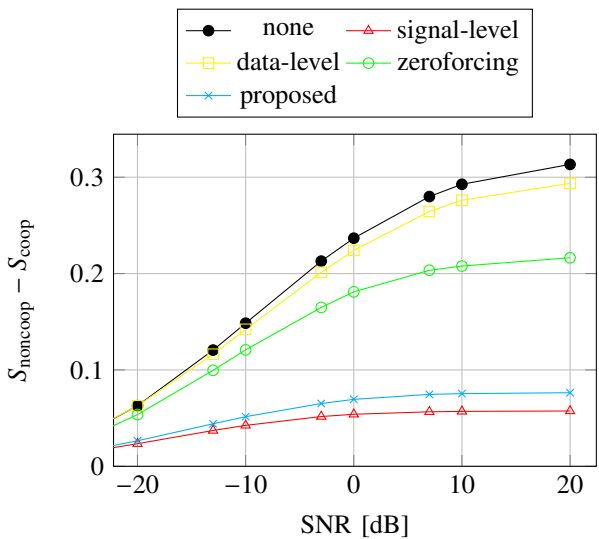

\subsection{Effects of Artificial Noise Design}
Existing approaches in the literature largely rely on transmitting uncorrelated Gaussian artificial noise, and can be broadly classified into two categories: 1) adding noise before signal scaling (e.g., \ac{dp}-achieving schemes) \cite{OtAFLDP,privacy-for-free, liaoOvertheAirFederatedLearning2022}, and 2) adding noise after scaling, leveraging the residual power not reserved for data transmission (e.g., schemes with friendly jammers) \cite{He2024secure,friendly-jammer,massny2023secure}. These are referred to as data- and signal-level noise, respectively, in the following results.\footnote{Direct comparison with existing secure \ac{ota} FL schemes is challenging due to different design objectives and performance metrics, cf. \cref{sec:related-works}.}
Fig. \ref{fig:Res2_noncoop} illustrates how the addition of artificial noise impacts the approximation error and security level in the non-cooperative case. Five different noise designs are implemented and compared:
\begin{itemize}
    \item \verb|none| : $\bm{A}=0$.
    \item \verb|signal-level| : Each user  independently transmits Gaussian noise with variance $P-\eta^2 / \abs{h_k}^2$.
    \item \verb|data-level| : Each user independently adds Gaussian noise prior to scaling the signal.
    \item \verb|zeroforcing| : Random zero-forcing design. $\bm{A}$ is chosen as a random $K\times(K-1)$ matrix that spans the entire nullspace of $\vect{h}$, multiplied by a scalar so that all rows satisfy their respective power constraints.
    \item \verb|proposed| : the optimized zero-forcing design computed according to (\ref{eq:proposed}).
\end{itemize}

All noise designs are simulated using the same $\eta$ value. The uncorrelated signal-level noise design achieves the highest security level since it leverages all available residual power as opposed to the other schemes. Conversely, not adding any artificial noise can achieve the lowest approximation error.

By design, zero-forcing approaches (including the proposed design) achieve the same approximation error as in the case without artificial noise. 
From a security perspective, the proposed method performs almost as well as the signal-level uncorrelated noise design, even though zero-forcing methods (and the data-level uncorrelated design) requires some users not to transmit with full power, to preserve the column span of $\bm{A}$ and fair aggregation, respectively.

In Fig. \ref{fig:Res2_coop}, we validate the performance of the proposed artificial noise design in the scenario with cooperating eavesdroppers. Similar to the non-cooperative case, the proposed design achieves security that approaches the case with signal-level uncorrelated noise. Additionally, it is worth noting that the proposed noise design closes the gap between the security levels in the non-cooperative and cooperative cases, respectively, which is further illustrated in Fig. \ref{fig:Res2_gap}. The advantage gained by the eavesdroppers through cooperation is significantly diminished under the proposed noise scheme. While cooperation remains a concern, the reduced security gap demonstrates that the proposed scheme provides robustness against eavesdropper cooperation, making it a less effective strategy for the eavesdroppers to exploit.

Furthermore, comparing the curves without artificial noise to those with signal- and data-level uncorrelated noise in Figs. \ref{fig:Res2_noncoop} and \ref{fig:Res2_coop}, it is evident that random perturbations disproportionately degrade the aggregation accuracy at the server. In the non-cooperative case (Fig. \ref{fig:Res2_noncoop}), this is due to the worst-case eavesdropper potentially experiencing a more favorable channel realization than the server. In the cooperative case (Fig. \ref{fig:Res2_coop}), eavesdroppers have access to multiple observations of the same perturbation signal, which can be exploited to obtain a more accurate estimate of the target function value.
This is the motivating observation behind the decision to optimize the noise precoding matrix across zero-forcing designs specifically. 

\subsection{Effects of Noisy Channel Estimates}
While the case of imperfect CSI was initially addressed in Section \ref{sec:nophase}, here we now consider a more realistic CSI uncertainty model. In this scenario, the legitimate users, the server, and the eavesdroppers have access to noisy channel estimates
\begin{align*}   
    \hat{\mathbf{h}} &= \mathbf{h} + \tilde{\mathbf{h}}, \\
    \hat{\mathbf{G}} &= \mathbf{G} + \tilde{\mathbf{G}}, 
\end{align*}
where $\tilde{\mathbf{h}} \sim \mathcal{CN}(\mathbf{0}, \sigma_h^2 \mathbf{I}_K)$ and $\text{vec}(\tilde{\mathbf{G}}) \sim \mathcal{CN}(\mathbf{0}, \sigma_G^2 \mathbf{I}_{LK})$ represent the estimation errors. These noisy estimates are utilized for artificial noise generation, signal encoding, and receive combining.
We assume that all parties have the same noisy channel estimates.
Fig. \ref{fig:noisy_channel_estimates} illustrates the empirical \ac{mse} at the server and eavesdroppers with the proposed artificial noise scheme for three distinct error levels: $\sigma_h^2 = \sigma_G^2 = 0$, $\sigma_h^2 = \sigma_G^2 = 10^{-5}$, and $\sigma_h^2 = \sigma_G^2 = 2 \times 10^{-5}$. 

It is observed that imperfect CSI significantly degrades the approximation performance at the server, while the eavesdropper \ac{mse} remains largely unaffected. This suggests that the security performance is relatively robust to the more realistic imperfect CSI setting considered here as well. 
Consistent with prior studies on \ac{ota} computation with imperfect CSI (cf., e.g., \cite{chen2022over}), the approximation error initially decreases with the \ac{snr} and subsequently increases due to the use of an estimator that does not account for channel estimation errors.

One explanation for the markedly different impact on the server and the eavesdroppers is that the server relies on coherent signal alignment and is therefore highly sensitive to the pre-processing mismatch caused by CSI errors. 
In contrast, the eavesdroppers inherently operate without channel alignment. This fundamental mismatch remains even when $\mathbf{h}$ is replaced by its estimate $\hat{\mathbf{h}}$ in the pre-processing step.

Nevertheless, CSI estimation errors can both exacerbate signal distortion at the eavesdroppers and reduce the effectiveness of the artificial noise, since the legitimate users align the artificial noise according to $\hat{\mathbf{G}}$ rather than the true channel matrix $\mathbf{G}$. In the considered setting, these effects appear to approximately cancel each other out. However, a rigorous analytical characterization of this phenomenon remains an interesting direction for future work.

\begin{figure}[!t]
\centering
\begin{tikzpicture}

\begin{axis}[
    xlabel={SNR [dB]},
    ylabel={Normalized MSE},
    xmode=log,
    log basis x=10, 
    xticklabel={$\pgfmathparse{10*\tick}\pgfmathprintnumber{\pgfmathresult}$},
    grid=major,
    width=7cm,
    height=5cm,
    every axis plot/.append style={line width=0.8pt},
    legend style={
        at={(0.5,1.05)},
        anchor=south,
        legend columns=2,
        cells={anchor=west},
    },
    mark options={solid},
    xminorticks=false,
]

\addplot[black, solid, mark=*, mark size=2pt]
table [x=P, y=D_zfo, col sep=space]
{tikz_data/noisy_channel_estimate_var_1e-05.dat};
\addlegendentry{D: $\sigma_h^2 = \sigma_G^2 = 0$}

\addplot[black, dashed, mark=*, mark size=2pt]
table [x=P, y=S1_zfo, col sep=space]
{tikz_data/noisy_channel_estimate_var_1e-05.dat};
\addlegendentry{S: $\sigma_h^2 = \sigma_G^2 = 0$}

\addplot[red, solid, mark=square, mark size=2pt]
table [x=P, y=D_zfo_noisy, col sep=space]
{tikz_data/noisy_channel_estimate_var_1e-05.dat};
\addlegendentry{D: $\sigma_h^2 = \sigma_G^2 = 10^{-5}$}

\addplot[red, dashed, mark=square, mark size=2pt]
table [x=P, y=S1_zfo_noisy, col sep=space]
{tikz_data/noisy_channel_estimate_var_1e-05.dat};
\addlegendentry{S: $\sigma_h^2 = \sigma_G^2 = 10^{-5}$}

\addplot[cyan, solid,mark=x]
table [x=P, y=D_zfo_noisy, col sep=space]
{tikz_data/noisy_channel_estimate_var_2e-05.dat};
\addlegendentry{D: $\sigma_h^2 = \sigma_G^2 = 2 \times 10^{-5}$}

\addplot[cyan, dashed,mark=x]
table [x=P, y=S1_zfo_noisy, col sep=space]
{tikz_data/noisy_channel_estimate_var_2e-05.dat};
\addlegendentry{S: $\sigma_h^2 = \sigma_G^2 = 2 \times 10^{-5}$}

\end{axis}
\end{tikzpicture}

\caption{Approximation error and security level of the proposed noise scheme under channel estimation errors with varying variances.}
\label{fig:noisy_channel_estimates}
\end{figure}

\subsection{Effects of Spatially Distributed Eavesdroppers}
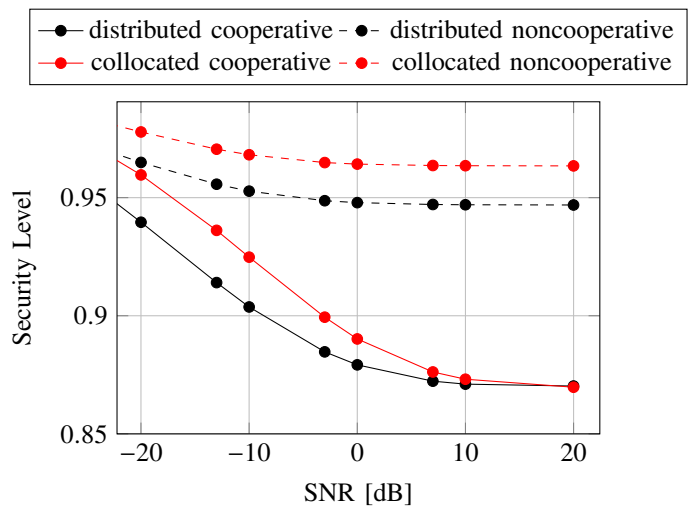
\begin{figure}[!t]
\centering
\begin{tikzpicture}[]

  \begin{axis}[
    xlabel={SNR [dB]},
    ylabel={Security Level},
    xticklabel={$\pgfmathparse{10*\tick}\pgfmathprintnumber{\pgfmathresult}$},
    xmode=log,
    log basis x=10,
    xminorticks=false,
    grid=major,
    ymin=0.85,
    width=7cm,
    height=5cm,
    every axis plot/.append style={line width=0.8pt},
    xmin=0.006,
    xmax=175,
    legend style={
      at={(0.5,1.05)},
      anchor=south,
      cells={align=left},
      legend columns=2
    },
    mark options={solid},
  ]

    \addplot[black, solid, mark=*, mark size=2pt] table [x=P, y=S1_dist, col sep=space] {tikz_data/collocated.dat};
    \addlegendentry{distributed cooperative}

    \addplot[black, dashed, mark=*, mark size=2pt] table [x=P, y=S2_dist, col sep=space] {tikz_data/collocated.dat};
    \addlegendentry{distributed noncooperative}

    \addplot[red, solid, mark=square, mark size=2pt] table [x=P, y=S1_collocated, col sep=space] {tikz_data/collocated.dat};
    \addlegendentry{collocated cooperative}

    \addplot[red, dashed,mark=square, mark size=2pt] table [x=P, y=S2_collocated, col sep=space] {tikz_data/collocated.dat};
    \addlegendentry{collocated noncooperative}
  \end{axis}
\end{tikzpicture}
\caption{Cooperative and non-cooperative security levels with collocated and distributed eavesdroppers.}
\label{fig:collocated}
\end{figure}

Some prior studies consider a single multiple-antenna eavesdropper \cite{server-jammer,luo2023joint,yao2025secure}. Figure \ref{fig:collocated} illustrates the security levels for the cooperative and non-cooperative cases with $L$ distributed single-antenna eavesdroppers compared to a single eavesdropper with $L$ antennas (collocating the eavesdroppers).
For a multi-antenna eavesdropper equipped with $L$ antennas, the channel matrix $\bm{G}$ is distributed such that its $k$-th column of follows $\vect{g}_k \sim \CN{0}{d_k^{-4} \bm{I}_L}$, where $d_k$ denotes the distance between the $k$-th user and the eavesdropper.
We observe that the distributed eavesdropper scenario considered in this paper constitutes a more challenging scenario in terms of ensuring a certain security level.

\subsection{Effects of Different Numbers of Zero-Forcing Users}
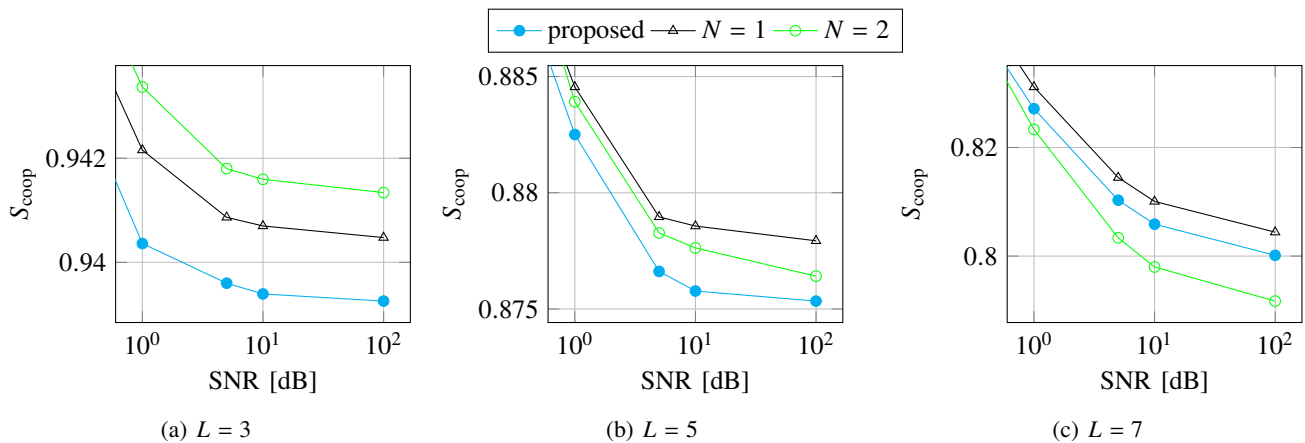
\begin{figure*}
\centering

\begin{subfigure}[t]{0.30\textwidth}
\centering
\begin{tikzpicture}
  \begin{axis}[
    xlabel={SNR [dB]},
    ylabel={$S_\text{coop}$},
    xmode=log,
    log basis x=10,
    xticklabel={$\pgfmathparse{10*\tick}\pgfmathprintnumber{\pgfmathresult}$},
    xminorticks=false,
    grid=major,
    yticklabel style={/pgf/number format/fixed, /pgf/number format/precision=3},
    xmin=0.6,
    width=5cm,
    height=4cm,
    every axis plot/.append style={line width=0.8pt},
    legend style={draw=none},
    mark options={solid},
  ]

    \addplot[cyan, solid, mark=*, mark size=2pt] table [x=P, y=proposed, col sep=space] {tikz_data/extension_L3.dat};
    
    \addplot[black, solid, mark=triangle*, mark size=3pt] table [x=P, y=N1, col sep=space] {tikz_data/extension_L3.dat};
    
    \addplot[green, solid, mark=square*, mark size=2pt] table [x=P, y=N2, col sep=space] {tikz_data/extension_L3.dat};

  \end{axis}
\end{tikzpicture}
\caption{$L=3$}
\end{subfigure}
\begin{subfigure}[t]{0.30\textwidth}
\centering
\begin{tikzpicture}
  \begin{axis}[
    xlabel={SNR [dB]},
    ylabel={$S_\text{coop}$},
    xmode=log,
    log basis x=10,
    xticklabel={$\pgfmathparse{10*\tick}\pgfmathprintnumber{\pgfmathresult}$},
    xminorticks=false,
    grid=major,
    yticklabel style={/pgf/number format/fixed, /pgf/number format/precision=3},
    xmin=0.6,
    width=5cm,
    height=4cm,
    every axis plot/.append style={line width=0.8pt},
    mark options={solid},
    legend style={
      at={(0.5,1.05)},
      anchor=south,
      legend columns=3
    },
  ]

    \addplot[cyan, solid, mark=*, mark size=2pt] table [x=P, y=proposed, col sep=space] {tikz_data/extension_L5.dat};
    \addlegendentry{proposed}
    
    \addplot[black, solid, mark=triangle*, mark size=3pt] table [x=P, y=N1, col sep=space] {tikz_data/extension_L5.dat};
    \addlegendentry{$N=1$}
    
    \addplot[green, solid, mark=square*, mark size=2pt] table [x=P, y=N2, col sep=space] {tikz_data/extension_L5.dat};
    \addlegendentry{$N=2$}
  \end{axis}
\end{tikzpicture}
\caption{$L=5$}
\end{subfigure}
\begin{subfigure}[t]{0.30\textwidth}
\centering
\begin{tikzpicture}
  \begin{axis}[
    xlabel={SNR [dB]},
    ylabel={$S_\text{coop}$},
    xmode=log,
    log basis x=10,
    xticklabel={$\pgfmathparse{10*\tick}\pgfmathprintnumber{\pgfmathresult}$},
    xminorticks=false,
    grid=major,
    xmin=0.6,
    width=5cm,
    height=4cm,
    every axis plot/.append style={line width=0.8pt},
    mark options={solid},
    legend style={draw=none},
  ]

    \addplot[cyan, solid, mark=*, mark size=2pt] table [x=P, y=proposed, col sep=space] {tikz_data/extension_L7.dat};
    \addplot[black, solid, mark=triangle*, mark size=3pt] table [x=P, y=N1, col sep=space] {tikz_data/extension_L7.dat};
    \addplot[green, solid, mark=square*, mark size=2pt] table [x=P, y=N2, col sep=space] {tikz_data/extension_L7.dat};

  \end{axis}
\end{tikzpicture}
\caption{$L=7$}
\end{subfigure}

\caption{Security level with different numbers of users ensuring zero-forcing in the proposed design, for different values of $L$.}
\label{fig:extension}
\end{figure*}

Fig. \ref{fig:extension} compares the average security levels with artificial noise computed according to (\ref{eq:proposed}) and (\ref{eq:extension}), respectively. 
The \verb|N=1| and \verb|N=2| curves are generated with optimal user selection found by exhaustive search. That is, for each channel realization, we solve (\ref{eq:extension}) for all possible subsets of $N$ users $\mathcal{Z}$, and choose the solution that gives the largest security level. Whereas, \verb|proposed| uses $\mathcal{Z} = \{ \argmax_k \abs{h_k}^2\}$. 
The small gap between the proposed and \verb|N=1| designs indicates that, if selecting a single user to ensure the zero-forcing condition, selecting the user with the best channel condition is close to optimal in most cases. 
The almost negligible security increase does not warrant the increased computational cost of constructing and evaluating multiple noise precoding matrices. 
When selecting multiple zero-forcing users, the gain depends on the number of eavesdroppers. For smaller values of $L$, where the eavesdropper channels span a smaller subspace, reducing the artificial noise's dimensionality (e.g., $K-N=8$ dimensions for \verb|N=2|) has a relatively small impact. In turn, it allows for an increased artificial noise amplitude, thus, improving the security level overall.
Conversely, for $L=7$, the increased magnitude of artificial noise generally fails to compensate for the reduction in security due to limiting the rank of $\bm{A}$. Nevertheless, it can be observed that $N=2$ outperforms $N=1$ for some channel realizations, suggesting that, even in this scenario, a scheme that adaptively chooses an optimal subset $\mathcal{Z}$ (including the value of $N$) can achieve even better performance. 
However, this requires an additional design to efficiently find the optimal user selection. In the remaining results, we only consider the case with one zero-forcing user, i.e., the design given in \eqref{eq:proposed}.

\subsection{Effects of Power Control}
\begin{figure}
\centering

\begin{tikzpicture}
  \begin{axis}[
    width=8cm,
    height=6cm,
    every axis plot/.append style={line width=0.8pt},
    xmin=0.006,
    xmax=175,
    xticklabel={$\pgfmathparse{10*\tick}\pgfmathprintnumber{\pgfmathresult}$},
    xmode=log,
    log basis x=10,
    xminorticks=false,
    grid=major,
    xlabel={SNR [dB]},
    ylabel={Normalized MSE},
    legend style={at={(0.45,1.05)}, anchor=south, legend columns=4, font=\scriptsize},
    mark options={solid},
  ]
  
  \addplot[green, dashed, mark=square*, mark size=2pt] table[x=P, y=S_0.4] {tikz_data/power_control_sim.dat};
  \addlegendentry{$S: \delta=0.40$}
  
  \addplot[blue, dashed, mark=triangle*, mark size=3pt] table[x=P, y=S_0.7] {tikz_data/power_control_sim.dat};
  \addlegendentry{$S: \delta=0.70$}

  \addplot[red, dashed, mark=*, mark size=2pt] table[x=P, y=S_0.85] {tikz_data/power_control_sim.dat};
  \addlegendentry{$S: \delta=0.85$}

  \addplot[black, dashed, mark=o, mark size=2pt] table[x=P, y=S_0.9999] {tikz_data/power_control_sim.dat};
  \addlegendentry{$S: \delta=1.0$}

  \addplot[green, solid, mark=square*, mark size=2pt] table[x=P, y=D_0.4] {tikz_data/power_control_sim.dat};
  \addlegendentry{$D: \delta=0.4$}
  
  \addplot[blue, solid, mark=triangle*, mark size=3pt] table[x=P, y=D_0.7] {tikz_data/power_control_sim.dat};
  \addlegendentry{$D: \delta=0.7$}

  \addplot[red, solid, mark=*, mark size=2pt] table[x=P, y=D_0.85] {tikz_data/power_control_sim.dat};
  \addlegendentry{$D: \delta=0.85$}

  \addplot[black, solid, mark=o, mark size=2pt] table[x=P, y=D_0.9999] {tikz_data/power_control_sim.dat};
  \addlegendentry{$D: \delta=1.0$}
  \end{axis}
\end{tikzpicture}
\caption{Security level and approximation error with the proposed noise design for different amplitude scaling factors.}
\label{fig:power_control_sim}
\end{figure}
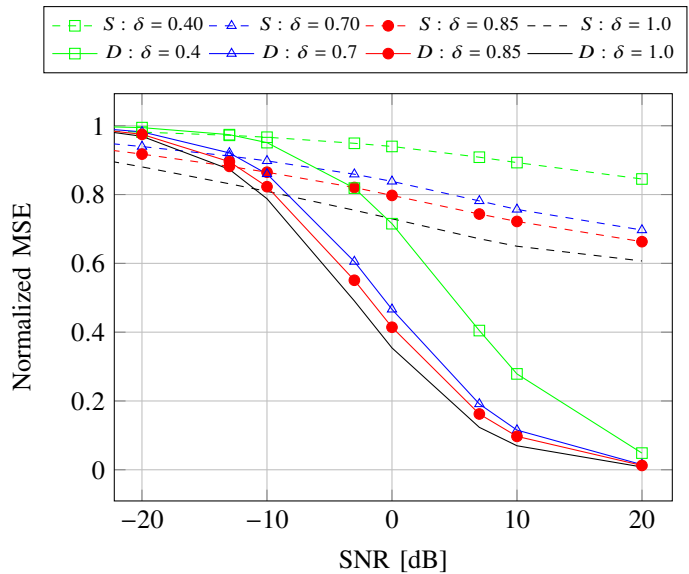

Fig. \ref{fig:power_control_sim} shows the impact of power control on the security level and approximation error, with the proposed noise design. 
For illustration purpose, we represent the amplitude scaling factor equivalently by the fraction $\delta \in [0,1]$ of the maximum value, that is, $\eta = \delta  \min_{k\in \{ 1,2,\ldots,K \} } P\abs{h_k}^2$ (cf. Corollary \ref{co:eta}).
At sufficiently high \ac{snr}, it is possible to greatly improve security (by decreasing $\eta$) without significantly impacting the approximation error.  

\subsection{Achievable Security-Accuracy Tradeoffs}
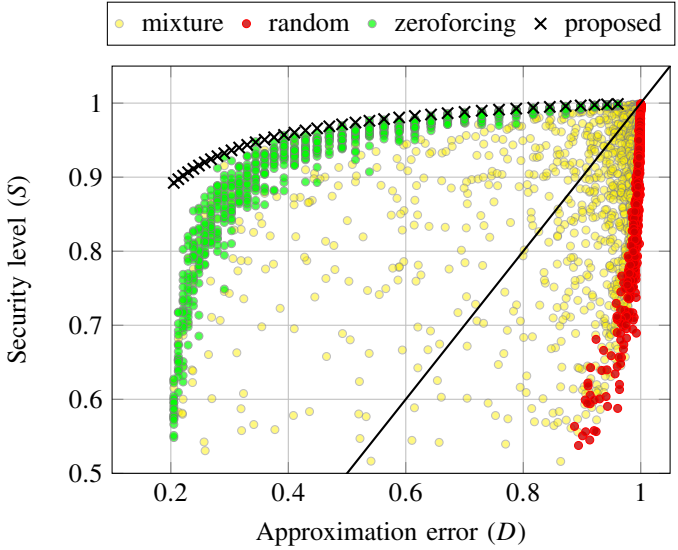
\begin{figure}
    \centering
    \begin{tikzpicture}
        \begin{axis}[
        width=8cm,
        height=6cm,
        xlabel={Approximation error ($D$)},
        ylabel={Security level ($S$)},
        grid=major,
        xmin=0.1, xmax=1.05, 
        ymin=0.5, ymax=1.05,
        legend style={
          at={(0.5,1.05)},
          anchor=south,
          legend columns=4,
          column sep = 5pt,
        },
        ]

        \addplot+[
        only marks,
        mark=square*,
        mark size=1.5pt,
        color=gray,
        opacity=0.5,
        mark options={solid, fill=orange}
        ] table[x=D, y=S] {tikz_data/tradeoff/mixing.dat};
        \addlegendentry{mixture}
        
        \addplot+[
        only marks,
        mark=*,
        mark size=1.5pt,
        fill = red,
        draw = none,
        fill opacity = 0.8,
        ] table[x=D, y=S] {tikz_data/tradeoff/random.dat};
        \addlegendentry{random}
        
        \addplot+[
        only marks,
        mark=triangle*,
        mark size=2.5pt,
        color=gray,
        mark options={solid, fill=green},
        opacity=0.7,
        ] table[x=D, y=S] {tikz_data/tradeoff/zf_random.dat};
        \addlegendentry{zeroforcing}
    
        \addplot+[
        only marks,
        mark=x,
        mark size=3pt,
        color=black,
        thick,
        ] table[x=D, y=S] {tikz_data/tradeoff/zf_opt.dat};
        \addlegendentry{proposed}
        
        \end{axis}
        \begin{axis}[
        width=8cm,
        height=6cm,
        xmin=0.1, xmax=1.05, 
        ymin=0.5, ymax=1.05,
        axis line style={draw=none},
        tick style={draw=none},
        ticks=none,
        ]
        \addplot[black, thick, domain=0:1.1, forget plot] {x};
        \end{axis}
        
    \end{tikzpicture}
    \caption{Achievable pairs of security level and approximation error for a given realization of the system.}
    \label{fig:achievable}
\end{figure}

 To better understand the security-accuracy trade-off, Fig. \ref{fig:achievable} numerically approximates the achievable pairs of security level and approximation error by Monte Carlo simulation. $L$ and the \ac{snr} are set to $7$ and $0 \, \text{dB}$, respectively.
 Only configurations above the diagonal are of potential interest, as they represent cases where the server achieves better estimation accuracy than the eavesdroppers. Noting that every artificial noise design can be decomposed into a zero-forcing part and a non-zero-forcing part, we generate random pairs of zero-forcing and non-zero-forcing noise designs and obtain mixtures by interpolating between them. This is repeated for $40$ equidistant values of $\delta$ in the range $[0,1]$.

It is possible that a random zero-forcing matrix achieves better security than the proposed scheme due to the large number of generated matrices. However, it is clear that the proposed design is among the top performers. 
Additionally, the proposed design can be found efficiently by solving a linear programming problem, which is preferred over generating and evaluating a large number of random designs in practice.
The proposed design performs particularly well when considering low approximation errors, or equivalently, high accuracy requirements at the server. In this regime, $\eta$ is large, leaving relatively little energy for artificial noise, necessitating careful use of the available power.

\section{Conclusions}
\label{sec:conclusion}
In this work, we investigated secure \ac{ota} computation in the presence of multiple (potentially cooperating) eavesdroppers by injecting correlated artificial noise at multiple source nodes to obscure their transmitted data signals. We provided fundamental bounds on the estimation performance of the server/receiver and the eavesdroppers, and derived the optimal joint estimator for cooperative eavesdroppers.
Furthermore, we proposed an optimized zero-forcing noise design, which was shown to achieve good performance under the accuracy-security tradeoff. Our main takeaway messages are as follows: 
1) In analog function computation over wireless channels, random phase rotation during signal transmission can provide inherent \ac{mse}-security against potential eavesdropping; 
2) This inherent security diminishes when multiple eavesdroppers are able to cooperate to improve their estimate of the intended computation result; 
3) Introducing correlated artificial noise with an optimized structure can substantially enhance security against eavesdropping without compromising the aggregation accuracy at the server.

{\appendices
\section{Proof of Theorem 1}
\label{app:th_approximation}
The approximation error is 
$$D =  \min_{d_y: \complexes \to \complexes} \frac{\EX{ \vert d_y(y;\bm{\Phi}) - s \vert ^2}}{\Var{s}},$$
where $s=\sum_{k=1}^K \gamma_k$ and $y = \eta s + \vect{h}^T\bm{A}\vect{v} + n_y$.
Since, $s$ and $y$ are jointly Gaussian, it holds that~\cite[Theorem 12.1]{kay2013fundamentals}
$$\EX{ \vert d_y(y;\bm{\Phi}) - s \vert ^2} = \Var{s} - \Cov{s,y}\Var{y}^{-1}\Cov{y,s},$$
and consequently 
$$D = 1 - \frac{\Cov{s,y}\Cov{y,s}}{\Var{y}\Var{s}}.$$
Since $n_y$ and all elements in $\bm{\gamma}$ and $\vect{v}$  are independent random variables with zero mean and unit variance, we have
\begin{align*}
    &\Var{s} = K, \\
    &\Var{y} = \Var{\eta s} + \Var{\vect{h}^T \bm{A} \vect{v}}+ \Var{n_y} \\ & \hspace{1cm} = \eta^2 K + \norm{\vect{h}^T \bm{A}}^2 + \sigma_y^2, \\
    &\Cov{s,y} = \Cov{y,s}\conj = \EX{\left(\eta s + \vect{h}^T\bm{A}\vect{v} + n_y \right)s\conj}
    \\ & \hspace{1cm} = \EX{\eta s s\conj} 
    = \eta K, 
\end{align*}
from which the theorem follows.
\hspace*{\fill} \qedsymbol 

\section{Proof of Theorem 2}
\label{app:th_security}
The \ac{mmse} estimator of $s$ is given by: 
$$\hat{s} = \EX{s| \vect{z}},$$
where $\vect{z}=(z_1,\ldots,z_L)^T$ contains the received signals at the eavesdroppers. Furthermore, since $s$ and $\vect{z}$ are jointly Gaussian, the \ac{mmse} estimator is linear and can be computed as~\cite[Theorem 12.1]{kay2013fundamentals}:
\begin{equation*}
    \hat{s} = \EX{s} + \Cov{s,\vect{z}}\Cov{\vect{z},\vect{z}}^{-1}\vect{z} = \EX{s\vect{z}^H}\EX{\vect{z}\vect{z}^H}^{-1}\vect{z}.
\end{equation*}
The corresponding \ac{mse} is 
\begin{align*}
    \Var{s\vert \vect{z}} &= \Var{s} - \Cov{s,\vect{z}}\Cov{\vect{z},\vect{z}}^{-1}\Cov{\vect{z},s} \\
    &= \Var{s} - \EX{s\vect{z}^H}\EX{\vect{z}\vect{z}^H}^{-1}\EX{\vect{z}s\conj},
\end{align*}
since both $s$ and $\vect{z}$ have zero mean. 
Given the expressions of $s$ and $\vect{z}$ written as
\begin{align*}
    &s =\vect{u}^T \bm{\gamma}, \\
     &\vect{z} = \eta \bm{G} \bm{D}_{\vect{h}}^{-1}\bm{\gamma} + \bm{G}\bm{A}\vect{v} + \vect{n}_z, 
\end{align*}
since all elements in $\bm{\gamma}$, $\vect{v}$ and $\vect{n}_z$ are independent random variables with zero mean and unit variance, we have
\begin{align*}
    &\Var{s} =K ,\\
    &\EX{s\vect{z}^H} = \eta \vect{u}^T \bm{D}_{\vect{h}}^{-H} \bm{G}^H ,\\
    &\EX{\vect{z} s\conj} = \eta\bm{G} \bm{D}_{\vect{h}}^{-1}\vect{u}, \\
    &\EX{\vect{z}\vect{z}^H} = \eta^2 \bm{G} \bm{D}_{\vect{h}}^{-1}\bm{D}_{\vect{h}}^{-H}\bm{G}^H  + \bm{G}\bm{A} \bm{A}^H\bm{G}^H + \sigma_z^2 \bm{I}_L.   
\end{align*}
From this, the security level can be derived, and we get 
\begin{equation*}
    \vect{p}_{\text{opt}} = \left( \EX{s\vect{z}^H}\EX{\vect{z}\vect{z}^H}^{-1} \right)^H = \EX{\vect{z}\vect{z}^H}^{-1}\EX{\vect{z} s\conj}.
\end{equation*}

\hspace*{\fill} \qedsymbol 

\bibliographystyle{IEEEtran}
\bibliography{references}

\end{document}